\documentclass[a4paper,11pt]{article}

\usepackage[a4paper,left=2.73cm,right=2.7cm,top=3cm,bottom=3.5cm]{geometry} 

\usepackage{amsmath,amssymb, mathrsfs}

\usepackage[colorlinks=true,linktocpage=true,linkcolor=blue,citecolor=blue,urlcolor=blue]{hyperref}

\usepackage[sort&compress,merge,numbers]{natbib}

\addtocontents{toc}{\protect\setcounter{tocdepth}{2}}
\numberwithin{equation}{section}

\RequirePackage[parfill]{parskip} 

\usepackage{epsfig}
\usepackage{graphicx}
\usepackage{epstopdf}
\usepackage{caption}
\usepackage{subcaption}

\def\wt{\widetilde}

\DeclareRobustCommand{\DIEP}{\ensuremath{%
\mathchoice{\includegraphics[height=2ex]{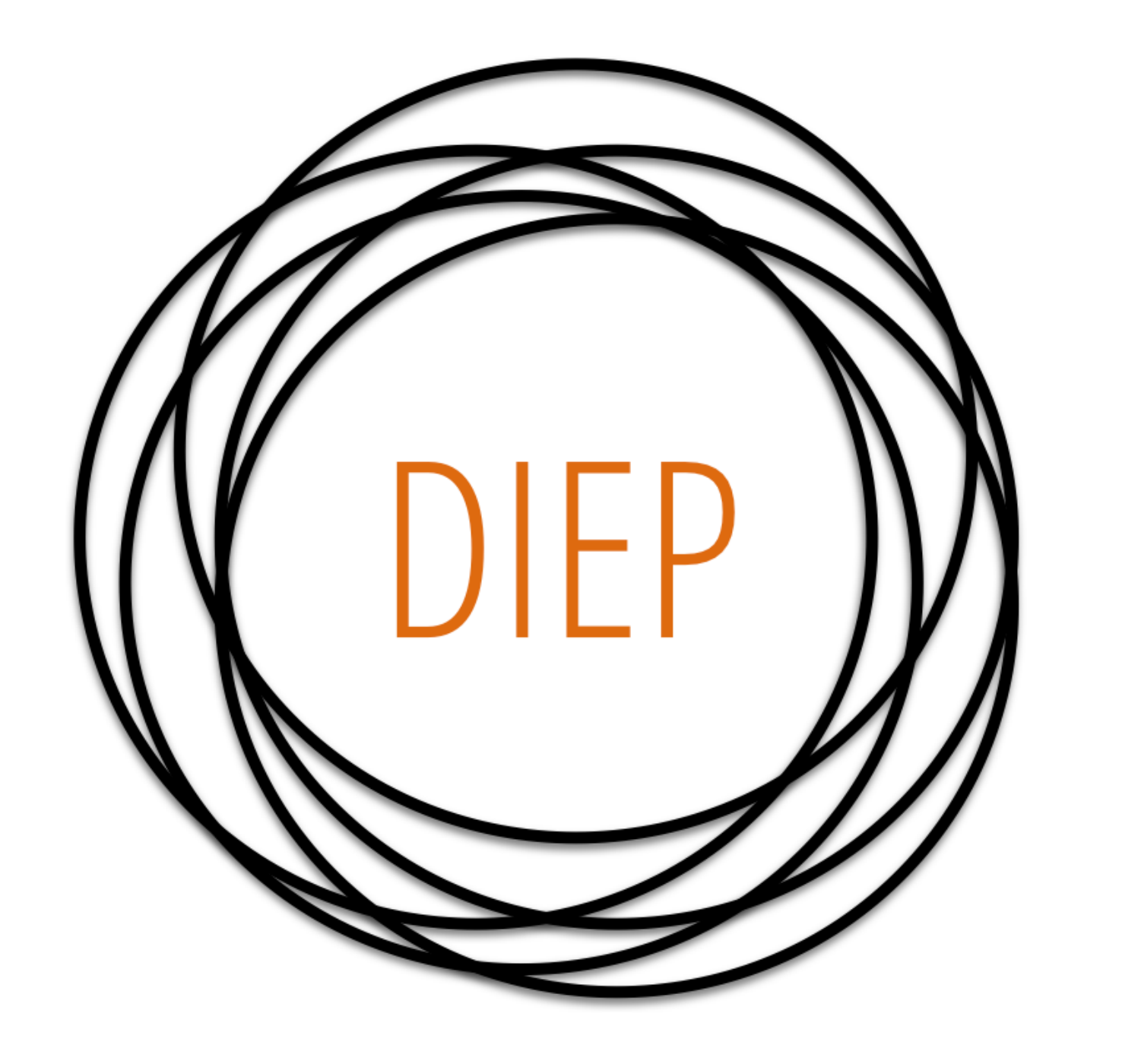}}
  {\includegraphics[height=2ex]{DIEPs.pdf}}
    {\includegraphics[height=1.5ex]{DIEPs.pdf}}
    {\includegraphics[height=1ex]{DIEPs.pdf}}
  }}

\usepackage[nostyling]{ajtex}                 

\begin{document}

\begin{titlepage}

\thispagestyle{empty}

\begin{flushright}
  \hfill{DCPT-18/01}
\end{flushright}

\vspace{40pt}  
	 
\begin{center}

{\Huge \textbf{Dissipative hydrodynamics }}
\vskip 0.3cm
{\Huge \textbf{with higher-form symmetry}}
	\vspace{30pt}
		
{\large \bf Jay Armas$^{1,\DIEP}$, Jakob Gath$^3$, Akash Jain$^4$ \\and Andreas Vigand Pedersen$^5$}
		
\vspace{25pt}

{$^1$Institute for Theoretical Physics, University of Amsterdam,\\
1090 GL Amsterdam, The Netherlands\\ 
\vspace{5pt}
$\DIEP$ Dutch Institute for Emergent Phenomena, The Netherlands \\
\vspace{5pt}
$^3$Department of Physics, Technical University of Denmark,\\
DK-2800 Kgs. Lyngby, Denmark \\
\vspace{5pt}
$^4$Centre for Particle Theory \& Department of Mathematical Sciences,\\
Durham University, South Road, Durham DH1 3LE, United Kingdom\\
\vspace{5pt}
$^5$Seaborg Technologies,\\
Bredgade 36B, 1260 Copenhagen, Denmark}

\vspace{20pt}
{\tt 
\href{mailto:j.armas@uva.nl}{j.armas@uva.nl}~,~\href{mailto:jagath@fysik.dtu.dk}{jagath@fysik.dtu.dk}~,\\\href{mailto:akash.jain@durham.ac.uk}{akash.jain@durham.ac.uk}~,~\href{mailto:andreas@seaborg.co}{andreas@seaborg.co}
}

\vspace{40pt}
				
\abstract{ A theory of parity-invariant dissipative fluids with $q$-form
  symmetry is formulated to first order in a derivative expansion. The fluid is
  anisotropic with symmetry $\text{SO}(D-1-q)\times\text{SO}(q)$ and carries
  dissolved $q$-dimensional charged objects that couple to a $(q+1)$-form
  background gauge field. The case $q=1$ for which the fluid carries string
  charge is related to magnetohydrodynamics in $D=4$ spacetime dimensions. We
  identify $q$+7 parity-even independent transport coefficients at first order
  in derivatives for $q>1$. In particular, compared to the $q=1$ case under the
  assumption of parity and charge conjugation invariance, fluids with $q>1$ are
  characterised by $q$ extra transport coefficients with the physical
  interpretation of shear viscosity in the $\text{SO}(q)$ sector and current
  resistivities.  We discuss certain issues related to the existence of a
  hydrostatic sector for fluids with higher-form symmetry for any $q\ge1$. We
  extend these results in order to include an interface separating different
  fluid phases and study the dispersion relation of capillary waves finding
  clear signatures of anisotropy. The formalism developed here can be easily
  adapted to study hydrodynamics with multiple higher-form symmetries.
 
}

\end{center}

\end{titlepage}

\tableofcontents

\hrulefill
\vspace{10pt}

\section{Introduction}\label{sec:intro}

The microscopic description of charged quantum matter is usually intractable
when the number of its fundamental objects is very large. Generically, however,
such microscopic descriptions admit a hydrodynamic limit in which the low-energy
collective behaviour of matter is captured by a few emergent degrees of freedom,
such as temperature, chemical potential and velocity fields. In this limiting
regime, rapidly varying quantities compared to the mean-free path of the
fundamental objects are integrated out while the dynamics of the slowly varying
quantities is governed by the conservation laws of the microscopic system. These
conservation laws are a direct manifestation of the underlying symmetries of the
system.

Despite hydrodynamics being a well established research subject, there has been
a substantial progress in its structural foundations in recent years. This
includes an off-shell formulation of hydrodynamics and the development of
classification schemes of its transport properties~\cite{Haehl:2014zda,
  Haehl:2015pja, Jain:2016rlz}; the construction of effective field theories and
the formulation of action principles for dissipative fluid
dynamics~\cite{Haehl:2015uoc, Crossley:2015evo, Glorioso:2017fpd,
  Glorioso:2017lcn, Jensen:2017kzi, Haehl:2017zac}; the
establishment of a framework for describing interfaces between different fluid
phases~\cite{Armas:2015ssd, Kovtun:2016lfw, Armas:2016xxg}; a new formalism for
studying non-relativistic fluids~\cite{Jensen:2014ama, Banerjee:2015hra,
  Jain:2015jla, Banerjee:2016qxf}; and the development of hydrodynamic theories with
generalised global 1-form symmetries and their connections to
magnetohydrodynamics~\cite{Schubring:2014iwa, Grozdanov:2016tdf,
  Hernandez:2017mch, Grozdanov:2017kyl} as well as their role in the
understanding of effective theories with translational symmetry breaking and
states with dynamical defects~\cite{Grozdanov:2018ewh}.

This paper introduces a framework for building effective hydrodynamic theories
of dissipative fluids with $q$-form symmetries, generalising previous work for
$q=0,1$. These effective theories correspond to the hydrodynamic limit of
microscopic descriptions whose underlying fundamental charged objects are
$q$-dimensional (i.e. $q$-dimensional branes). These $q$-dimensional objects
couple to a background gauge field $A_{q+1}$. In the language of
\cite{Gaiotto:2014kfa}, these fluids describe microscopic systems with a
generalised $q$-form global symmetry. Associated with the $q$-form symmetry is a
$(q+1)$-form current $J$ whose integral over a $(D-q-1)$ dimensional
hypersurface ${\mathcal{M}_{\Gamma}}$ yields a conserved dipole charge
\begin{equation}
Q_{{\mathcal{M}_{\Gamma}}}=\int_{\mathcal{M}_{\Gamma}}\star J~~,
\end{equation}
where the operator $\star$ is the Hodge dual operator in $D$-dimensional
spacetime. This dipole charge counts the number of $q$-dimensional objects that
cross the $(D-q-1)$-dimensional hypersurface
${\mathcal{M}_{\Gamma}}$.\footnote{In the case $q=1$ in $D=4$ and in the context
  of magnetohydrodynamics where $J_{\mu\nu} = \star F_{\mu\nu}$, the fundamental
  objects are strings and $Q_{{\mathcal{M}_{\Gamma}}}$ counts the number of
  magnetic field lines crossing a codimension-2
  hypersurface. See~\cite{Grozdanov:2016tdf} for a detailed discussion.} The
hydrodynamic theories constructed here capture the collective excitations of
these charged $q$-dimensional objects around a state of thermal equilibrium.

This work has been highly motivated by the structure of long-wavelength
perturbations of black branes in supergravity, where dissipative fluids with
higher-form symmetries are naturally realised~\cite{Emparan:2011hg,
  Caldarelli:2010xz, Emparan:2013ila, DiDato:2015dia, Armas:2016mes}. As the
fundamental fields in supergravity include several higher-form gauge fields,
generic black brane bound states can carry multiple higher-form (dipole)
charges. For instance, the D3-F1 bound state in type IIB string theory carries
two 2-form currents and one 4-form current~\cite{Armas:2016mes}.  Within this
context, the fluid stress tensor and charge currents appear as the effective
currents sourcing the gravitational and electric fields far away from the black
brane horizon~\cite{Armas:2016mes}, while their conservation laws are realised
as constraint equations when solving Einstein equations in a long-wavelength
expansion.

Tackling the general problem of establishing a hydrodynamic theory of
dissipative fluids carrying multiple higher-form charges is certainly of
interest and in this work we take the first step towards this goal by
constructing the effective hydrodynamic theory of fluids with one single $q$-form
symmetry. These fluids are anisotropic with $\text{SO}(D-q-1)\times\text{SO}(q)$
symmetry and the source of this anisotropy is the existence of a higher-form
charge current. Gravitational duals to these fluids are encountered in black
brane geometries in gravity theories with a metric $g_{\mu\nu}$, one single
$(q+1)$-form gauge field $A_{q+1}$ and possibly a dilaton field. Several
examples of such gravitational duals with arbitrary $q$ were found in
\cite{Caldarelli:2010xz}.

In this work, we formulate the theory of dissipative fluids with $q$-form
symmetry to first order in a long-wavelength expansion (to first order in
derivatives of the fluid fields) focusing on the parity-even sector of the
theory.\footnote{The $q=1$ case deserves considerably more attention and will be the
  focus of a later publication~\cite{toappear2}. In this paper we further
  restrict the $q=1$ case by requiring charge conjugation invariance in addition
  to parity invariance. In this case, our analysis for $q=1$ is the same as that inpro~\cite{Grozdanov:2016tdf}.} We also study equilibrium
configurations and highlight some of the technical complications that arise
while describing the hydrostatic sector of fluids with higher-form
symmetries. In addition, we generalise our results to include the presence of
interfaces separating different fluid phases and study specific cases of surface
waves, finding signatures of anisotropy in their dispersion relation.


\subsection{Summary of the results and organisation of the material}

In Sec.~\ref{sec:ideal}, the ideal order dynamics of fluids with $q$-form
symmetry living in a background spacetime with metric $g_{\mu\nu}$ and a
$(q+1)$-form gauge field $A_{q+1}$ is introduced. These fluids are anisotropic
with $\text{SO}(D-q-1)\times\text{SO}(q)$ symmetry and carry a dipole charge
density $Q$. They are characterised by a stress tensor, charge $(q+1)$-form
current and an entropy current. The existence of this $(q+1)$-form current is
responsible for the microscopic anisotropic properties of the fluid. When the
charge $Q$ vanishes, the current vanishes as well and the usual uncharged
isotropic fluid is recovered. We establish their thermodynamic properties and
conservation laws. As far as we are aware, this is the first time that the ideal
order dynamics of these fluids is formalised in the literature.

In Sec.~\ref{sec:diss} we formulate the dissipative sector of the theory up to
first order in derivatives. We focus on the parity-even sector for $q\ge0$ while
for $q=1$ we in addition require charge conjugation invariance. First, we
describe how frame transformations act on the stress tensor and currents and
comment on different choices of frames. Picking a higher-form analogue of the
Landau frame, we proceed and require the divergence of the entropy current to be
positive semi-definite. This leads to the existence of $q$+8 transport
coefficients for $q>1$ and 8 for $q=1$, all of which are dissipative, thereby
contributing to entropy production. Once Kubo formulae are obtained, we note
that Onsager's relation for mixed correlation functions sets a constraint among
these transport coefficients, thereby leading to $q$+7 independent transport
coefficients at first order in derivatives for $q>1$ and 7 for $q=1$. Compared
to the $q=1$ case studied in~\cite{Grozdanov:2016tdf, Hernandez:2017mch}, for
$q>1$ there is one additional transport coefficient with the physical
interpretation of shear viscosity in the $\text{SO}(q)$ sector and $q-1$ extra
current resistivities. At the end of this section, we study the constraints on
these transport coefficients in isotropic limits. We observe that some of these
constraints are satisfied by gravitational duals even away from the isotropic
limits.

Sec.~\ref{sec:equilibrium} contains a detailed analysis of equilibrium
configurations in theories with a $q$-form symmetry. We begin by noting that the
equilibrium partition function presented in~\cite{Grozdanov:2016tdf} for $q=1$
does not describe the hydrostatic sector of the theory completely. As such, the
hydrostatic solution as defined in~\cite{Grozdanov:2016tdf}, which assumes
hydrostatic backgrounds to admit not just a timelike isometry but also $q$
spacelike isometries, causes production of entropy, which is incompatible
with equilibrium. We show
that to avoid this, an additional constraint must be imposed on the hydrostatic
backgrounds. Furthermore,~\cite{Grozdanov:2016tdf} assumed that spacelike and
timelike isometries admitted by the hydrostatic backgrounds must have vanishing
mutual inner products, which further restricts the class of backgrounds on which
equilibrium can be realised. We explicitly derive the most general partition
function where these inner products are not assumed to be zero 
and show that the resulting solution
is a $q$-form generalisation of the $q=1$ solution presented
in~\cite{Caldarelli:2010xz}. We comment on other issues regarding the
hydrostatic sector of these theories, wherein the requirement of an equilibrium
partition function seems to impose more constraints than requiring the second
law of thermodynamics alone to hold. This is in striking contrast with $q=0$
hydrodynamics, where the second law guarantees the existence of an equilibrium
partition function at all orders in the derivative
expansion~\cite{Bhattacharyya:2013lha, Bhattacharyya:2014bha}.

In Sec.~\ref{sec:surface}, following~\cite{Armas:2015ssd, Armas:2016xxg}, we
generalise our results in order to include the presence of an interface/surface
separating two different fluid phases. Similar to Sec.~\ref{sec:equilibrium}, we
write down a partition function for fluids with $q$-form symmetry in the
presence of the interface. We then analyse the divergence of the surface entropy
current and obtain the surface thermodynamics as well as a constraint on the
normal component of the bulk fluid velocity. We observe that this matches
partition function expectations. Having established a notion of equilibrium in
this setting, we obtain the dispersion relation for capillary waves and ripples
on the interface, finding clear signals of anisotropy. 

Finally, in Sec.~\ref{sec:discussion} we comment on some open issues and future
research directions. We also provide App.~\ref{app:frames} with some of our
results written in another fluid frame in order to ease comparison with earlier
literature.

\section{Ideal order fluids with $q$-form symmetry}\label{sec:ideal}

In this section we introduce the ideal order currents and conservation equations
for the propagation of an anisotropic fluid with $q$-form symmetry carrying
$q$-brane charge in a $D$-dimensional background geometry
$\left(\mathcal{M},g_{\mu\nu}, A_{q+1}\right)$ with $p$-spatial directions so
that $D=(p+1)$ with $p\ge q$. The manifold $\mathcal{M}$ is endowed with the
Levi-Civita connection $\nabla_\mu$ built out from the background metric
$g_{\mu\nu}$ with coordinates $x^{\mu}$. These fluids are characterised by a
$(q+1)$-form current $J$ that couples to the background gauge field
$A_{q+1}$. In general, introducing a conserved higher-form current breaks the
$\text{SO}(p)$ symmetry enjoyed by the ordinary relativistic ``point charged''
(or neutral) fluid to a $\text{SO}(p-q)\times\text{SO}(q)$ symmetry.\footnote{If
  multiple $q_i$-form conserved currents with $i=1,\ldots,\ell$ are introduced,
  the fluid is expected to have the symmetry
  $\text{SO}(p-\sum_{i=1}^\ell q_i )\times \text{SO}(q_i)\times\cdots\times
  \text{SO}(q_\ell)$.} As usual, at each point of $\mathcal{M}$ there exists a
rest-frame in which the fluid is static. At ideal order, this frame is
unambiguously defined and is characterised by a timelike vector $u^{\mu}$ (the
fluid velocity) normalised such that $u^\mu u_\mu=-1$. For later use, we introduce
the projector transverse to the fluid velocity
$\Delta^{\mu\nu} = g^{\mu\nu} + u^{\mu}u^{\nu}$.
The local thermodynamic fields of the fluid are then unambiguously defined as
their local values in the rest-frame. We now proceed to write now the ideal
order hydrodynamics describing this system.


\subsection{Fluid stress tensor and current}\label{sec:fluidst}
At the ideal order level, the $(q+1)$-form current takes the following form
\begin{equation}
  \label{eq:cur0}
  J_{\text{ideal}} = Q~\text{Vol}_{(q+1)}~~.
\end{equation}
Here $Q$ denotes the quasi-local dipole charge and $\text{Vol}_{(q+1)}$ is a
$(q+1)$-dimensional volume form. We can decompose it as
$\text{Vol}_{(q+1)}\equiv u\wedge \text{Vol}_{q}$, where
$\text{Vol}_{q}=-\imath_u \text{Vol}_{(q+1)}$ (here $\imath$ denotes the
interior product) is the local volume form on the space spanned by the $q$
(linearly independent) spatial directions $v^{i}$ of the current in the
rest-frame, i.e.
\begin{equation}
  \label{eq:volq}
  \text{Vol}_{q}=\sqrt{\text{det}\left(v_i\cdot v_j\right)}~
  v^{1}\wedge\ldots\wedge v^{q}~~, \qquad
  v^{i}_{\mu}u^{\mu} = 0~~.
\end{equation}
Here we have used the inverse of the matrix $\left(v^i\cdot v^j\right)$ defined as
$g^{\mu\nu}v^i_\mu v^j_\nu$, which is the induced metric on the $q$-manifold. We will
justify this picture and notation below. Clearly, $\text{Vol}_{q}$ and $u^{\mu}$
are both $\text{SO}(q)$ invariant structures. Moreover, $\text{Vol}_{(q+1)}$ is
a $\text{SO}(1,q)$ Lorentz invariant under boosts along the $q$-directions
$v^{i}$ with $i=1,\ldots,q$. This invariance must be reflected in the
hydrodynamic theory once dissipation is introduced.

In general, the current induces stresses along the $q$-spatial directions of the current. We thus introduce a $\text{SO}(q)$ invariant projector $\Pi_{\mu\nu}$ along these directions. In particular, this projector satisfies the relations
\begin{equation}
  {\Pi^{\mu}}_\rho{\Pi^{\rho}}_\nu = {\Pi^{\mu}}_\nu~~,\qquad
  {\Pi^{\mu}}_\rho u^{\rho} = 0~~,\qquad
  \text{Tr}\Pi\equiv g^{\mu\nu}\Pi_{\mu\nu}=q~~.
\end{equation}
Without loss of generality, one may choose the $q$ one-forms $v^i$ to be
orthonormal, leading to the projector\footnote{\label{symmetry_footnote} Note
  that $\text{Vol}_{q}$ is invariant under a local $GL(q)$ transformation:
  $v^{i}_{\mu} \to \tensor{R}{^{i}_{j}} v^{j}_{\mu}$, where
  $\tensor{R}{^{i}_{j}}$ is an arbitrary $q\times q$ non-singular matrix. By
  choosing to work in an orthonormal basis for the vectors $v^{i}_{\mu}$, we have reduced
  this $GL(q)$ symmetry down to a residual $SO(q)$.}
\begin{equation}
  \Pi_{\mu\nu} = \delta_{ij} v^{i}_\mu v^{j}_\nu~~,\qquad
  g^{\mu\nu} v^{i}_\mu v^{j}_\nu = \delta^{ij} ~~,
\end{equation}
thereby justifying the form of~\eqref{eq:volq}. Given this projector, we
introduce the ideal order fluid stress tensor as
\begin{equation}
  \label{eq:stress0}
  T^{\mu\nu}_{\text{ideal}} = E ~ u^{\mu}u^{\nu} + P_q~\Pi^{\mu\nu} + P~\Gamma^{\mu\nu}~~,
\end{equation}
where $\Gamma^{\mu\nu}\equiv \Delta^{\mu\nu} - \Pi^{\mu\nu}$ is the
$\text{SO}(p-q)$ invariant projector orthogonal to the current. This projector
satisfies $\Gamma^{\mu\rho}\Pi_{\rho\nu}=\Gamma^{\mu\rho}u_{\rho}=0$ and
$\text{Tr}\Gamma=(p-q)$. We also find it convenient to introduce the projector
onto all timelike and spacelike directions of the charge current,
$\Xi^{\mu\nu} = - u^{\mu} u^{\nu} + \Pi^{\mu\nu}$. The stress tensor
\eqref{eq:stress0} is the most general stress tensor compatible with the given
symmetries. In~\eqref{eq:stress0} we have introduced the energy density
$E$ such that $u_\nu T^{\mu\nu}_{\text{ideal}} = - E u^\mu$ while $P_q$
and $P$ denote the pressure along and orthogonal to the current, respectively.

The stress tensor also satisfies the orthogonality condition
$\Pi_{\mu\rho}\Gamma_{\nu\sigma}T^{\mu\nu}_{\text{ideal}}=0$, which will play an
important role in the choice of fluid frame when discussing dissipative
effects. Note that, in particular, when $q=0$ (equivalently $\Pi_{\mu\nu}=0$) or
when $q=p$ (equivalently $\Gamma_{\mu\nu}=0$) we recover the complete
$\text{SO}(p)$ isotropy as required.\footnote{The dynamics of $p=q$ fluids has
  been extensively analysed in~\cite{Emparan:2011hg, DiDato:2015dia}.} Isotropy
is also recovered in the uncharged (or neutral) limit ($Q=0$) since the source
of anisotropy is only due to the existence of the current~\eqref{eq:cur0}. In
turn, this implies that $(P-P_q)|_{Q=0}=0$.
Requiring consistency with the existence of an uncharged isotropic limit will
impose non-trivial constraints among transport coefficients in the dissipative
sector of the theory in that limit.

                    
\subsection{Thermodynamics}
                    
The local thermodynamics of the $q$-charged fluid is analogous to the local
thermodynamics of the $q=1$ fluid first considered
in~\cite{Caldarelli:2010xz}. In particular, the fluid energy density satisfies
the first law of thermodynamics
\begin{equation}
  \label{eq:t1}
  \text{d}E = T\text{d}S + \mu \text{d}Q~~,
\end{equation} 
with $T$, $S$ and $\mu$ denoting the temperature, entropy density and chemical
potential dual to $Q$, respectively.\footnote{These thermodynamic properties are
  motivated from black brane geometries~\cite{Emparan:2011hg, Caldarelli:2010xz,
    Armas:2016mes} but we also show in Sec.~\ref{sec:equilibrium} that they
  naturally follow from partition function considerations.} We assume that the
fluid carries an entropy current of the form
\begin{equation}
  \label{eq:ent0}
  S_{\text{ideal}}^{\mu} = S u^{\mu}~~,
\end{equation}
which we require to obey a local form of the second law of thermodynamics
$\nabla_\mu S^{\mu}_{\text{ideal}}\ge0$.

In addition, the difference between the pressure orthogonal and along the current is given by the energy density of the dissolved $q$-branes carried by the fluid, i.e.
\begin{equation}
  \label{eq:rel}
  P-P_q=\mu Q~~.
\end{equation}
The local thermodynamics of the fluid with $q$-brane charge also satisfy the Gibbs-Duhem relations
\begin{equation}
  \label{eq:gd}
  E + P = T S+\mu Q~~,\qquad
  \text{d}P = S\text{d}T+Q\text{d}\mu~~,\qquad
  \text{d}P_q = S \text{d} T-\mu \text{d} Q~~.
\end{equation}
One may easily see that relation~\eqref{eq:rel} is the result of integrating the last two relations in~\eqref{eq:gd}. Furthermore, it is clear that in the uncharged limit ($Q=0$) one obtains the condition $\left(P-P_q\right)|_{Q=0}=0$.


\subsection{Conservation equations}

The stress tensor and current obey the usual conservation laws
\begin{equation}\label{eq:claws}
  \nabla_\mu T^{\mu\nu}
  = F^\nu~~, \qquad
  \text{d}\star J = 0~~,
\end{equation}
where $F^\nu \equiv \frac{1}{(q+1)!} F^{\nu\mu_1\ldots\mu_{q+1}}
J_{\mu_1\ldots\mu_{q+1}}$ is a Lorentz force acting on the fluid and where the
$(q+2)$-form $F$ is the field strength $F=\text{d}A_{q+1}$. Here $\star$ is the
Hodge dual on $\mathcal{M}$. We first consider the charge conservation
equation. Taking the wedge product of $\text{d}\star J = 0$ with $J$ itself we obtain
\begin{equation}
  \star J \wedge \star \left(\text{d}\star J \right)=\star \mathfrak{f}~~,
\end{equation}
where at ideal order, $\mathfrak{f}$ is the one-form
\begin{equation}
  \mathfrak{f}_\mu= {(-1)}^{q}Q~ \Xi_{\mu\lambda}\nabla_\nu\left(Q ~ \Xi^{\nu\lambda}\right)~~.
\end{equation}
By projecting out this equation along the $u_{\mu}$ and $\Pi_{\mu\nu}$
directions, the vanishing of the one-form $\mathfrak{f}$ then implies two
continuity equations for the charge $Q$, namely
\begin{equation}\label{eq:current0}
  u^{\mu}\nabla_\mu Q = - Q ~ \vartheta_{(p-q)}~~,~~
  {\Pi_\mu}^{\nu}\nabla_\nu Q
  = Q~ \Pi_{\mu}{}^{\nu} \left(\mathfrak{a}_\nu
    - \nabla_\lambda\Pi^{\lambda}{}_{\nu} \right)~~,
\end{equation}
where we have defined the fluid expansion
$\vartheta_{(p-q)}=\Gamma^{\mu\nu}\nabla_\mu u_{\nu}$ in the $\text{SO}(p-q)$
sector as well as the fluid acceleration
$\mathfrak{a}^{\mu}=u^{\lambda}\nabla_\lambda u^\mu$.
We now look at the charge conservation equation in directions orthogonal to the
current. Let us introduce $(p-q)$ one-forms $\gamma^{a}_{\mu}$ orthogonal to
$u^{\mu}$ and $\Pi^{\mu\nu}$ and mutually orthonormal. Contracting the
conservation equation with $\gamma^{a}_{\mu}$, we obtain
\begin{equation} \label{eq:perp}
  \Xi^{\lambda\mu}{(\text{d}\gamma^{a})}_{\mu\nu}\Xi^{\nu\rho}=0~~.
\end{equation}
By virtue of Frobenius' theorem, the set of one-forms $\gamma^a$ are thus
surface-forming, that is, in each point there exists an integral
$(q+1)$-dimensional submanifold of the vectors $u$ and $v^i$. It follows that
$\mathcal{M}$ is foliated into a set of $(q+1)$-dimensional submanifolds which
can essentially be thought of as the level-surfaces for the dipole charge
$Q$. Furthermore, the induced metric on these submanifolds is $\Xi_{\mu\nu}$ and
the volume-form is precisely the structure we previously denoted by
$\text{Vol}_{(q+1)}$ in~\eqref{eq:cur0}-\eqref{eq:volq}. By projecting
\eqref{eq:perp} along $u$ and $v^i$ one finds the set of equations
\begin{equation} \label{eq:commuting}
 {\Gamma}^{\nu}{}_{\lambda} \left(v_i^\mu\nabla_\mu v^\lambda_{j} - v_j^\mu \nabla_\mu
   v^\lambda_{i} \right)
 = {\Gamma}^{\nu}{}_{\lambda} \left(u^\mu\nabla_\mu v^\lambda_{i} - v_i^\mu
   \nabla_\mu u^\lambda\right) = 0~~.
\end{equation}
These equations will become important when analysing equilibrium configurations.

We now consider the conservation of the stress tensor~\eqref{eq:claws} at ideal
order and project it along the $u^{\mu}$, $\Pi^{\mu\nu}$ and $\Gamma^{\mu\nu}$
directions, obtaining the three equations of motion
\begin{equation} \label{eq:stcon}
  \begin{split}
    u^{\mu}\nabla_\mu E
    &= - (E+ P) \nabla_{\mu}u^{\mu}
    + \mu Q ~ \vartheta_q ~~, \\
    (E+P)\Pi_{\mu}{}^{\nu}\mathfrak{a}_\nu
    &= \mu Q ~ \Pi_{\mu}{}^{\nu}\nabla_{\lambda}\Pi^{\lambda}{}_{\nu}
    -{\Pi_\mu}^{\nu}\nabla_\nu (P-\mu Q)~~,\\
    (E+P){\Gamma_\mu}^{\nu}\mathfrak{a}_\nu
    &= \mu Q ~{\Gamma_{\mu}}^{\nu}\nabla_\lambda \Pi^{\lambda}{}_{\nu}
    - {\Gamma_{\mu}}^{\nu}\nabla_\nu P+{\Gamma_{\mu\nu}} F^{\nu}~~,
  \end{split}
\end{equation}
where we have defined the fluid expansion $\vartheta_q=\Pi^{\mu\nu}\nabla_\mu u_{\nu}$
in the $\text{SO}(q)$ sector.
Using the charge conservation equations~\eqref{eq:current0} and the
thermodynamic properties~\eqref{eq:t1}-\eqref{eq:gd}, we may rewrite the stress
conservation equations~\eqref{eq:stcon} as the conservation of entropy current
and Euler force equations\footnote{The last equation in \eqref{eq:stcon1} can
  also be written as
  \begin{equation}\nonumber
    S T ~{\Gamma_\mu}^{\nu}\left(\mathfrak{a}_\nu+ \nabla_\nu \ln T\right)
    - Q\mu ~ {\Gamma_\mu}^{\nu} \left(\mathcal{K}_{\nu}
      - \nabla_\nu \ln \mu\right)
    = {\Gamma_{\mu\nu}} F^{\nu},
  \end{equation}
  where $\mathcal{K}^\mu$ is the mean extrinsic curvature of the $q$-brane
  embedded into the $(p+1)$-dimensional space, defined as
  $\mathcal{K}^\mu=\Xi^{\nu\lambda}\nabla_\nu{\Xi_\lambda}^\mu$. This is a
  higher-form generalisation of the analogous equation derived for $q=1$ and
  $F^{\mu}=0$ in~\cite{Caldarelli:2010xz}.}
\begin{equation}\label{eq:stcon1}
  \begin{split}
    \nabla_\mu\left(Su^\mu\right)
    &= 0~~,\\
    \Pi_{\mu}{}^{\nu}\lb \mathfrak{a}_{\nu}
    + \frac1T\nabla_\nu T \rb
    &= 0~~,\\
    (E+P) ~{\Gamma_\mu}^{\nu}\left(\mathfrak{a}_\nu
      + \frac1T \nabla_\nu T\right)
    + Q T ~ {\Gamma_\mu}^{\nu} \left(
      \nabla_\nu \bfrac{\mu}{T}
      - \frac{\mu}{T} \nabla_{\lambda}\Pi^{\lambda}{}_{\nu}
      - \frac{1}{T} E_{\nu} \right)
    &= 0 ~~,
  \end{split}
\end{equation}
where
$E_\mu= \frac{1}{q!} u^{\mu_1}\text{Vol}_q^{\mu_2\ldots\mu_{q+1}}F_{\mu
  \mu_1\ldots\mu_q+1}$.

It is instructive to perform an explicit counting of independent degrees of freedom
and the number of dynamical equations determining the time evolution of a given fluid configuration.
There are $p$ independent components of the fluid velocity $u^{\mu}$, as one component is fixed
by the normalisation condition $u^\mu u_\mu=-1$, and one degree of freedom associated with
the temperature $T$. Corresponding to these $(p+1)$ degrees of freedom are $(p+1)$ dynamical
equations provided by the stress tensor conservation equation \eqref{eq:stcon1}. In addition, there are $(p-q)q$ independent components of
$v^{i}_{\mu}$
and one degree of freedom associated with the chemical potential $\mu$.\footnote{The $q$ one-forms have
$q(p+1)$ components but there are $q(q+1)/2$ mutual orthonormality
conditions, $q$ vanishing inner products with $u^{\mu}$ and $q(q-1)/2$ components
which are not independent due to $\SO(q)$ symmetry. This leads to $(p-q)q$ independent components.}
Their dynamics is provided by the charge conservation equations with a time-derivative, i.e. 
the first equation in \cref{eq:current0} and the second equation in
\cref{eq:commuting} with $(p-q)q$ components. Therefore there is a match between the number of
independent degrees of freedom and the number of dynamical equations. This continues to hold
at higher order in derivatives since, as we will show below, it is always possible to choose a 
fluid frame for which the corrections to the current $\delta J$ satisfy $\imath_u \delta J=0$. The remaining equations in
\eqref{eq:current0} and \eqref{eq:commuting} provide consistency requirements for the initial
conditions on a Cauchy slice.

We will return to all these conservation equations in Sec.~\ref{sec:equilibrium}
where we discuss equilibrium configurations. In the following section we
formulate the theory at first order in derivatives.


\section{Dissipative fluids with $q$-form symmetry}\label{sec:diss}

Having defined the hydrodynamics of an ideal anisotropic fluid, we now explain
how to include derivative corrections. Here we shall follow the approach to
relativistic dissipative hydrodynamics originally introduced by Landau and
Lifshitz~\cite{Landau:1987gn}. This approach entails postulating the existence
of an entropy current $S^{\mu}$ which to any given order is constructed from the
available hydrodynamic operators. The entropy current is then constrained
on-shell by an ultra-local version of the second law of thermodynamics, namely,
$\nabla_{\mu}S^{\mu}\ge0$. In this section we first comment on possible choices
of fluid frames and then, after picking a higher-form generalisation of the
Landau frame, impose the second law of thermodynamics leading to $q$+8 transport
coefficients for $q>1$ and $8$ transport coefficients for $q=1$. This is
followed by a derivation of Kubo formulae, which, when combined with Onsager's
relation further constrains the transport coefficients, reducing the total
number of transport coefficients by one. Finally, we study the constraints on the transport
coefficients in different isotropic limits discussed in Sec.~\ref{sec:fluidst}.


\subsection{Dissipative corrections and choices of fluid frames}

In a derivative expansion, all the hydrodynamic currents are corrected in powers
of the expansion parameter such that the total stress tensor $T^{\mu\nu}$,
charge current $J^\mu$ and entropy current $S^{\mu}$ can be written as
\begin{equation}
  T^{\mu\nu} = T^{\mu\nu}_{\text{ideal}}
  + \delta T^{\mu\nu} + \mathcal{O}(\partial^2)~,\quad
  J = J_{\text{ideal}}
  + \delta J + \mathcal{O}(\partial^2)~,\quad
  S^{\mu} = S^{\mu}_{\text{ideal}}
  + \delta S^{\mu} + \mathcal{O}(\partial^2)~,
\end{equation}
where $\delta T^{\mu\nu}$, $\delta J$ and $\delta S^{\mu}$ are
$\mathcal{O}(\partial)$ derivative corrections whose general form we seek to
find. We decompose the corrections to these currents according to the available
symmetries
\begin{equation} \label{eq:exp}
  \begin{split}
    \delta T^{\mu\nu}
    &= \alpha~ u^{\mu} u^{\nu}
    + 2\left(u^{(\mu}\phi_{\Pi}^{\nu)}
      + u^{(\mu}\phi_{\Gamma}^{\nu)}\right)
    + \tau_{\Pi}\Pi^{\mu\nu}+\tau_{\Gamma}\Gamma^{\mu\nu}
    + \tau^{\mu\nu}~~, \\
    \delta J
    &= \beta ~\text{Vol}_{(q+1)}+\psi\wedge \text{Vol}_q+\Upsilon~~, \\
    \delta S^{\mu}
    &= \frac1T \phi_\Pi^{\mu} + \frac{1}{T} \phi_\Gamma^{\mu}
    - \frac{\mu}{T} \psi^{\mu}
    + \delta S^{\mu}_{\text{non-can}}~~.
\end{split}
\end{equation}
In~\eqref{eq:exp} we have introduced the first order vector $\phi_{\Pi}^{\mu}$
describing the heat flux along the $\Pi^{\mu\nu}$ directions, subject to the
constraints $u_{\mu} \phi_{\Pi}^{\mu} = \Gamma_{\mu\nu} \phi_\Pi^\nu =
0$. Similarly, $\phi_\Gamma^\mu$ is a heat flux vector along $\Gamma^{\mu\nu}$
satisfying $u_{\mu} \phi_{\Pi}^{\mu} = \Pi_{\mu\nu} \phi_\Gamma^\nu = 0$. We
have also introduced $\tau_{\Pi} = \Pi_{\mu\nu}\delta T^{\mu\nu}$ and
$\tau_\Gamma = \Gamma_{\mu\nu}\delta T^{\mu\nu}$ which denote the trace of
$\delta T^{\mu\nu}$ along the $\Pi_{\mu\nu}$ and $\Gamma_{\mu\nu}$ directions
respectively. Furthermore, $\tau^{\mu\nu}$ is a symmetric and traceless tensor
subject to the constraints
$u_\mu\tau^{\mu\nu} = \Pi_{\mu\nu}\tau^{\mu\nu} =
\Gamma_{\mu\nu}\tau^{\mu\nu}=0$. In the decomposition of the charge current, we
introduced the one-form $\psi_{\mu}$ such that
$u^{\mu} \psi_{\mu} = \Pi^{\mu\nu}\psi_\nu=0$. In addition, $\Upsilon$ is a
$(q+1)$-form orthogonal to $\delta J$, that is,
$\text{Vol}_q\wedge \star\Upsilon=0$. The scalars $\alpha$ and $\beta$ are
composed of linear combinations of the one-derivative hydrodynamic scalars
available, namely $\vartheta_{(p-q)}$ and $\vartheta_q$.

The first three terms in the entropy current are made out of vector structures
which are already present in the stress tensor and charge current, and are
commonly referred to as the \emph{canonical entropy current}. The last term
$\delta S^{\mu}_{\text{non-can}}$ is referred to as the \emph{non-canonical entropy current}
 and accounts for independent tensor structures that can appear in the
entropy current. As we shall see, up to first order in derivatives and when restricted to the
parity-even sector, $\delta S^{\mu}_{\text{non-can}}$ is forced to vanish due to the second law of
thermodynamics.\footnote{For $q=1$ we further require
  charge conjugation invariance, implying that terms appearing in the
  constitutive relations must be invariant under the transformation
  $v^\mu_1\to-v^\mu_1$. Relaxing parity and charge-conjugation invariance leads
  to many more transport coefficients. In the case of $q=1$, many of these have
  been written down in~\cite{Hernandez:2017mch}. In a future publication, we
  will revisit this case in further detail~\cite{toappear2}.} Therefore the
entropy current at first order in derivatives takes the canonical form. Nevertheless,
we include this term here in order to facilitate the analysis of the 
hydrostatic sector of the theory.

\subsubsection{Fluid frames}\label{sec:frames}

So far, the discussion has been rather general and the decomposition and
constraints have been imposed on symmetry grounds and to match the anticipated
number of degrees of freedom. However, as usual, one must specify a fluid frame
due to the freedom of redefining the fluid variables $T$, $\mu$, $u^{\mu}$,
$v^i_{\mu}$ according to
\begin{equation}
  T\to T+\overline\delta T~~,\quad
  \mu\to\mu+\overline\delta\mu~~,\quad
  u\to u^{\mu} + \overline\delta u^{\mu} ~~,\quad
  v^i\to v^i_{\mu} +\overline\delta v^i_{\mu} ~~,
\end{equation}
subjected to the orthogonality conditions
\begin{equation}
  u_{\mu}\overline\delta u^{\mu} = 0~~,\quad
  u^{\mu} \overline\delta v^i_{\mu}
  + v^{i}_{\mu} \overline\delta u^{\mu} =0~~,\quad
  v^{i\mu} \overline\delta v^j_{\mu}
  + v^{j\mu} \overline\delta v^i_{\mu}
  = 0 ~~\forall~~i,j=1,\ldots,q~~.
\end{equation}
This frame transformation leads to the following change in the stress tensor and charge current
\begin{equation} \label{eq:framet}
  \begin{split}
    \overline\delta T^{\mu\nu}
    &=  \overline\delta E ~ u^{\mu}u^{\nu}
    +  \overline\delta P_q ~ \Pi^{\mu\nu}
    +  \overline\delta P ~ \Gamma^{\mu\nu}
    + 2(E + P) u^{(\mu} \overline\delta u^{\nu)}
    - 2\mu Q ~  \delta^{ij} v^{(\mu}_i \overline\delta v^{\nu)}_j~~,\\
    \overline\delta J
    &=  \overline\delta Q ~ \text{Vol}_{(q+1)}
    + Q ~  \overline\delta u\wedge \text{Vol}_q+Qu\wedge\sum_{i=1}^q \left(\imath_{v^i}\text{Vol}_q\right)\wedge  \overline\delta v^i~~.
\end{split}
\end{equation}
For a fluid with particle charge ($q=0$) a convenient choice of frame is the
Landau frame defined by the conditions
$u_\mu \delta T^{\mu\nu} = u_\mu \delta J^{\mu}=0$. For arbitrary $p$ and $q$,
the obvious generalisation of the Landau conditions, which reduces to the
Landau frame in the isotropic limits ($q=0$ or $q=p$ or $Q=0$), is defined by
\begin{equation} \label{eq:Landau}
  u_\mu \delta T^{\mu\nu}=0~~,~~
  \Pi_{\mu\lambda} \Gamma_{\nu\rho} \delta T^{\mu\nu}=0~~,~~
  \star\left(\star \text{Vol}_{(q+1)}\wedge \delta J\right)=0~~.
\end{equation}  
This choice of frame implies that $\alpha=\beta=\phi_{\Pi}=\phi_{\Gamma}=0$ and that $\tau_{\mu\nu}$ satisfies the orthogonality condition $\Pi_{\mu\lambda} \Gamma_{\nu\rho} \tau^{\mu\nu}=0$. 

Another common frame used in the context of dissipative fluids with particle charge is the Eckart frame, defined as $u_\mu u_\nu\delta T^{\mu\nu}= \delta J^\mu=0$. We note that in the general case of a fluid charged under a $(q+1)$-form current, there is no direct analogue of the Eckart frame in the sense that it is not possible to set $\delta J=0$ since there is no frame transformation~\eqref{eq:framet} that can eliminate the orthogonal components of the current ${\Gamma_{\nu_1}}^{\mu_1}\ldots{\Gamma_{\nu_{q+1}}}^{\mu_{q+1}}J^{\nu_1\ldots\nu_{q+1}}$. The closest analogue of the Eckart frame is defined by the conditions
\begin{equation}
  u_\mu u_\nu\delta T^{\mu\nu}=0~~,\qquad
  u_\mu\Pi_{\nu\lambda}\delta T^{\mu\lambda} = 0 ~~,\qquad
  \star\left(\text{Vol}_{q}\wedge \star\delta J\right)=0~~,\qquad
  \imath_u \delta J=0~~.
\end{equation}
This frame implies that $\alpha=\beta=\phi_{\Pi}=\psi=0$ and the constraint
$\imath_u \Upsilon=0$. This in turn has the consequence that the interpretation given
by Eq.~\eqref{eq:perp} of $\mathcal{M}$ being foliated into a set of
$(q+1)$-dimensional submanifolds is in general lost, though it is recovered in
the hydrostatic sector, as we shall see in Sec.~\ref{sec:equilibrium}. In the
core of this paper we have choosen to use the Landau frame~\eqref{eq:Landau} but
in App.~\ref{app:frames} we present the results of the next section in the frame
of~\cite{Grozdanov:2016tdf, Hernandez:2017mch}, which is different than the
frame~\eqref{eq:Landau}.


\subsection{Entropy current constraints}

We are now ready to obtain the constraints following from the second law of
thermodynamics. To begin with, we evaluate the divergence of the entropy current
given in~\eqref{eq:exp} without making reference to a particular hydrodynamic
frame. After a bit of algebra, we obtain
\begin{align}\label{eq:divSframeless}
  \nabla_{\mu}S^{\mu}
  =& - \frac{1}{T^{2}} \alpha ~ u^{\mu}\nabla_{\mu}T
     - \frac1T \lb \phi_{\Pi}^{\mu} + \phi_{\Gamma}^{\mu} \rb
     \lb \frac{1}{T}\nabla_{\mu}T + \mathfrak{a}_\mu \rb\nn\\
   & - \frac{\tau_\Gamma}{T} \vartheta_{(p-q)}
     - \frac{\tau_\Pi}{T}\vartheta_q
     - \frac{1}{T}\tau^{\mu\nu}\nabla_\mu u_\nu \nn\\
   & - \beta~ u^{\mu}\nabla_{\mu}\bfrac{\mu}{T}
     - \psi^{\mu} \left(\dow_{\mu}\bfrac{\mu}{T}
    - \frac{\mu}{T}\nabla_\nu{\Pi^{\nu}}_\mu
    - \frac{1}{T}E_{\mu} \right)
     - \frac{1}{T}\star\Big[\star\Upsilon\wedge\left(
     \mu\,\text{dVol}_q + \imath_u F\right)\Big] \nn\\
   &+ \nabla_{\mu}\delta S^{\mu}_{\text{non-can}} ~~,
\end{align}
where we have defined the interior product of $(q+2)$-form field strength
as ${(\imath_u F)}_{\mu_{1}\ldots\mu_{q+1}} = u^\mu
F_{\mu\mu_1\ldots\mu_{q+1}}$.  Given our choice of fluid
frame~\eqref{eq:Landau}, these expressions can be simplified to
\begin{align} \label{eq:divS}
  \nabla_{\mu}S^{\mu}
  &= - \psi^{\mu} \left(\dow_{\mu}\bfrac{\mu}{T}
    - \frac{\mu}{T}\nabla_\nu{\Pi^{\nu}}_\mu
    - \frac{1}{T}E_{\mu} \right)
     - \frac{1}{T}\star\Big[\star\Upsilon\wedge\left(
     \mu\,\text{dVol}_q + \imath_u F\right)\Big] \nn\\
   &\qquad -\frac{1}{T}\tau^{\mu\nu}\nabla_\mu
     u_\nu-\frac{\tau_\Gamma}{T} \vartheta_{(p-q)}
     - \frac{\tau_\Pi}{T}\vartheta_q
     + \nabla_{\mu}\delta S^{\mu}_{\text{non-can}} ~~.
\end{align}
Our main task now is to make this expression manifestly positive semi-definite
for any hydrodynamic configuration. We begin by focusing on the non-canonical
piece. Note that every term other than
$\nabla_{\mu}\delta S^{\mu}_{\text{non-can}}$ is a product of two one-derivative
tensor structures. Therefore, to ensure positive definiteness, we must ensure
that $\nabla_{\mu}\delta S^{\mu}_{\text{non-can}}$ does not contain any pure two
derivative terms. By an explicit counting, it is possible to show that
no such terms can appear in $\delta S^{\mu}_{\text{non-can}}$
when imposing parity-invariance (and charge-conjugation invariance for
$q=1$). Thus, $\delta S^{\mu}_{\text{non-can}}$ vanishes at first
order in derivatives. We now analyse the remaining terms, proceeding term-by-term.

The first term in~\eqref{eq:divS} implies that
\begin{equation}
  \psi_\mu=-\mathfrak{D} T h_{\{\mu\}}~~,\qquad
  \{h\} = h_{\{\mu\}}dx^\mu =
  {\Gamma_\mu}^{\nu}\left(\nabla_\nu\left(\frac{\mu}{T}\right)
    - \frac{\mu}{T}\nabla_\lambda{\Pi^\lambda}_\nu
    - \frac{1}{T}E_\nu\right)dx^\mu~~,
\end{equation}
where the function $\mathfrak{D}(T,\mu)$ satisfies $\mathfrak{D}(T,\mu)\ge0$ and where curly brackets denote projection onto $\Gamma_{\mu\nu}$. The vector $\psi^\mu$ is the usual charge diffusion vector with an associated diffusion constant $\mathfrak{D}$ (Fick's diffusion law). As expected, the diffusion flux is only sensitive to the thermodynamic charges orthogonal to the current. This means that the diffusion vector vanishes in the case $q=p$, as it should~\cite{Gath:2013qya, DiDato:2015dia}.

We now consider the second term in~\eqref{eq:divS}, which leads to transport
coefficients that have no analogue in fluids carrying particle charge. The
simplest way to make this term positive semi-definite is by taking
$\Upsilon\sim \mu\,\text{dVol}_q + \imath_u F$, however, this is not consistent
with the constraint $\text{Vol}_q\wedge \star\Upsilon=0$. We therefore need to
project $\mu\,\text{dVol}_q + \imath_u F$ against $\text{Vol}_{q}$. Since
$\mu\,\text{dVol}_q + \imath_u F$ is a $(q+1)$-form as opposed to
$\text{Vol}_{q}$ which is a $q$-form, we need to project at least two indices of
$\mu\,\text{dVol}_q + \imath_u F$ along the $u^{\mu}$ or $\Gamma^{\mu\nu}$
directions. Let us first define
\begin{equation}\label{VolqU}
  \begin{split}
    {\left(\text{dVol}_q^\Gamma\right)}_{\mu_1\ldots\mu_{q+1}}
    &\equiv {\Gamma^{\mu}}_{[\mu_1}{\Gamma^{\nu}}_{\mu_2}
    {\left(\text{dVol}_q\right)}_{|\mu\nu|\mu_3\ldots\mu_{q+1}]}~~, \\
    {\left(\text{dVol}_q^U\right)}_{\mu_1\ldots\mu_{q+1}}
    &\equiv {u^{\mu}}u_{[\mu_1}{\Gamma^{\nu}}_{\mu_2}
    {\left(\text{dVol}_q\right)}_{|\mu\nu|\mu_3\ldots\mu_{q+1}]}~~.
  \end{split}
\end{equation}
Note that due to the definition of $\text{Vol}_{q}$ in~\eqref{eq:volq},
contracting more than two of its indices along $u^{\mu}$ or $\Gamma^{\mu\nu}$
leads to zero. In light of this,~\eqref{VolqU} are the most generic contractions
of $\text{dVol}_q$ along $u^{\mu}$ and $\Gamma^{\mu\nu}$. Since the
contraction of $\imath_{u}F$ with $u^{\mu}$ vanishes, the projection of
$\mu\,\text{dVol}_q + \imath_u F$ along one $u^{\mu}$ and one $\Gamma^{\mu\nu}$
direction leads to $\mu\,\text{dVol}_q^U$. The remaining independent projections
on the other hand are
\begin{equation}
  \mu\,\text{dVol}_{q}^{\Gamma} + \Gamma^{2}(\imath_{u}F)~~, ~~
  \Gamma^{I+1}(\imath_{u}F) ~~\text{for}~~ 2\leq I\leq q~~,
\end{equation}
where $\Gamma^{I+1}(\imath_{u}F)$ has $I+1$ of its indices contracted along
$\Gamma^{\mu\nu}$ and the remaining along $\Pi^{\mu\nu}$. The number of such
projections is bounded by $p-q$, i.e.\ the number of independent $\Gamma^{\mu\nu}$
directions available. Having isolated the tensor structures, we can write down
the most generic expression for $\Upsilon$ as
\begin{equation}
  \Upsilon=
  - \Theta ~\text{dVol}_q^U
  - \Omega_{1}\left(\mu\,\text{dVol}_q^\Gamma
    + \Gamma^{2}(\imath_u F)\right)
  - \sum_{I=2}^{q}\Omega_{I} \Gamma^{I+1}(\imath_{u} F)~~,
\end{equation}
where $\Theta(T,\mu)$ and $\Omega_{I}(T,\mu)$ satisfy
$\Theta(T,\mu),\Omega_{I}(T,\mu)\ge0$ and are new types of current resistivities
that appear for fluids carrying $q$-brane charge.

We now look at the corrections due to non-trivial terms in the stress
tensor. We start by looking at the term involving $\tau_{\mu\nu}$. To this end we decompose the covariant derivative of the fluid velocity according to
 \begin{equation} \label{eq:dev}
\nabla_\mu u_\nu= - u_\mu \mathfrak{a}_\nu +\omega_{\mu\nu}+\frac{\vartheta_q}{q}\Pi_{\mu\nu}+\frac{\vartheta_{(p-q)}}{p-q}\Gamma_{\mu\nu}+\Sigma_{\mu\nu}~~.
\end{equation}
Here
$\omega_{\mu\nu}={\Delta_\mu}^{\lambda}{\Delta_\nu}^{\rho}\nabla_{[\lambda}u_{\rho]}$
is the vorticity of the fluid while $\Sigma_{\mu\nu}$ is the symmetric and
traceless part of $\nabla_\mu u_\nu$ orthogonal to $u$ decomposed according to
the available $\text{SO}$-sectors such that\footnote{For the purposes of
  comparison, note that for an isotropic fluid the decomposition~\eqref{eq:dev}
  may be expressed as
  \begin{equation}\nonumber
    \nabla_\mu u_\nu=-u_\mu \mathfrak{a}_\nu
    + \omega_{\mu\nu} +\frac{\vartheta_p}{p}\Delta_{\mu\nu}
    + \frac{1}{2}{\Delta_\mu}^{\lambda}{\Delta_\nu}^{\rho}\left(\mathscr{L}_u g_{\lambda\rho}-\frac{1}{p}g_{\lambda\rho}\text{Tr}\mathscr{L}_u g\right)~~.
 \end{equation}}
\begin{equation} \label{eq:sigma}
    \Sigma_{\mu\nu}=\frac{1}{2}\left(\mathscr{L}_u \Gamma_{\{\mu\nu\}}
      + \mathscr{L}_u \Pi_{\langle\mu\nu\rangle}\right)
    + \mathscr{L}_u \Gamma_{\{(\mu\nu)\rangle}
    + \mathscr{L}_u \Pi_{\{(\mu\nu)\rangle}~~.
  \end{equation}
Here $\mathscr{L}_u$ denotes the Lie derivative with respect to $u^{\mu}$ and the
curly (angled) brackets denote projection onto $\Gamma_{\mu\nu}$
($\Pi_{\mu\nu}$) and subtraction of the trace. For example
\begin{equation}
  \mathscr{L}_u \Gamma_{\{\mu\nu\}}
  = {\Gamma^{\lambda}}_\mu{\Gamma^{\rho}}_\nu
  \left(u^\alpha\nabla_\alpha\Gamma_{\lambda\rho}
    + \Gamma_{\alpha\rho}\nabla_\lambda u^\alpha
    + \Gamma_{\lambda\alpha}\nabla_\rho u^\alpha
    - \frac{2}{p-q}\Gamma_{\lambda\rho}
    \Gamma^{\alpha\beta}\nabla_{\alpha}u_{\beta}\right)~~.
\end{equation}
Using~\eqref{eq:sigma}, we see that it is possible to add three distinct terms
to $\tau_{\mu\nu}$: $\mathscr{L}_{u}\Pi_{\langle\mu\nu\rangle}$,
$\mathscr{L}_{u}\Gamma_{\lbr\mu\nu\rbr}$ and
$\mathscr{L}_u \Gamma_{\{(\mu\nu)\rangle}+\mathscr{L}_u
\Pi_{\{(\mu\nu)\rangle}$. Note that via projections of $\nabla_{\mu}u_{\nu}$ only the
linear combination of $\mathscr{L}_u \Gamma_{\{(\mu\nu)\rangle}$ and
$\mathscr{L}_u \Pi_{\{(\mu\nu)\rangle}$ provides an independent tensor structure. Therefore $\tau_{\mu\nu}$ can only depend on
this sum. In any case, the second Landau condition in~\eqref{eq:Landau}
eliminates a possible
$\mathscr{L}_u \Gamma_{\{(\mu\nu)\rangle}+\mathscr{L}_u \Pi_{\{(\mu\nu)\rangle}$
term in $\tau^{\mu\nu}$. This implies that in this frame there is no shear
cross-viscosity between the two sectors. Therefore, we are lead to take
\begin{equation}
\tau_{\mu\nu}=-\eta_{\Pi}\mathscr{L}_u \Pi_{\langle\mu\nu\rangle}-\eta_{\Gamma}\mathscr{L}_\mu \Gamma_{\{\mu\nu\}}~~,
\end{equation}
where $\eta_{\Pi}(T,\mu)$ and $\eta_{\Gamma}(T,\mu)$ satisfy $\eta_{\Pi}(T,\mu),\eta_{\Gamma}(T,\mu)\ge0$ and denote the shear viscosity in the $\text{SO}(q)$ and $\text{SO}(p-q)$ sectors, respectively. 

Finally, we look at the $\tau_{\Pi}$ and $\tau_{\Gamma}$ terms
in \eqref{eq:divS}. The most general ansatz for these terms is given in terms of first
order scalars such that
\begin{equation}
  \tau_\Pi = - q\left(\zeta_\Pi ~ \vartheta_q
    + \zeta_{\Pi,\Gamma}~\vartheta_{(p-q)} \right)~~,~~
  \tau_\Gamma = -(p-q)\left(\zeta_{\Gamma}~ \vartheta_{(p-q)}
    + \zeta_{\Gamma,\Pi}~ \vartheta_q\right)~~.
\end{equation}
Here $\zeta_\Pi$ is a bulk viscosity in the $\text{SO}(q)$ sector, $\zeta_{\Gamma}$ a bulk viscosity in the $\text{SO}(p-q)$ sector and $\zeta_{\Pi,\Gamma},\zeta_{\Gamma,\Pi}$ are two bulk cross-viscosities. This ansatz, when introduced into~\eqref{eq:divS} leads to the following contribution to the divergence of the entropy currrent
\begin{equation}
  \nabla_{\mu}S^{\mu} \sim
  \frac{1}{T}
  \begin{pmatrix}
    \vartheta_{(p-q)} \\
    \vartheta_q
  \end{pmatrix}^{\mathrm T}
  \begin{pmatrix}
    (p-q)\zeta_{\Gamma} & (p-q)\zeta_{\Gamma,\Pi} \\
     q~\zeta_{\Pi,\Gamma} & q~\zeta_\Pi
  \end{pmatrix}
  \begin{pmatrix}
    \vartheta_{(p-q)} \\
    \vartheta_q
  \end{pmatrix}~~.
\end{equation}
Requiring this quadratic form to be positive semi-definite implies that we must have 
\begin{equation}
  \zeta_{\Gamma}\ge0~~,\qquad
  \zeta_\Pi\ge0~~,\qquad
  q(p-q)\zeta_{\Gamma}\zeta_\Pi \ge \frac{1}{4}{\left((p-q)\zeta_{\Gamma,\Pi}
    + q\zeta_{\Pi,\Gamma}\right)}^2~~.
\end{equation}
This completes the requirements of the second law of thermodynamics. 

Summarising, we have the following first order derivative corrections to the stress tensor, charge and entropy current\footnote{As earlier advertised, the total entropy current up to first order in
derivatives takes the canonical form
\begin{equation} \nonumber
S^{\mu}=s u^\mu-\frac{1}{T}\delta T^{\mu\nu}u_\nu-\frac{\mu}{T}\delta J^{\mu \mu_1\ldots\mu_{q}}\left({\text{Vol}_q}\right)_{\mu_1\ldots\mu_q}~~.
\end{equation}}
\begin{align} \label{eq:1storder}
  \delta T_{\mu\nu}
  &= - \eta_{\Pi}\mathscr{L}_u \Pi_{\langle\mu\nu\rangle}
  - \eta_{\Gamma}\mathscr{L}_u \Gamma_{\{\mu\nu\}} \nn
- \Pi_{\mu\nu}\left(\zeta_\Pi\vartheta_q +
    \zeta_{\Pi,\Gamma}\vartheta_{(p-q)}\right)
  - \Gamma_{\mu\nu}\left(\zeta_{\Gamma} \vartheta_{(p-q)}
    + \zeta_{\Gamma,\Pi}\vartheta_q\right)~~, \nn\\
  \delta J
  &= -\mathfrak{D} T \{h\}\wedge \text{Vol}_q
    - \Theta ~\text{dVol}_q^U
    - \Omega_{1}\left(\mu\,\text{dVol}_q^\Gamma
      + \Gamma^{2}(\imath_u F)\right)
    - \sum_{I=2}^{q}\Omega_{I} \Gamma^{I+1}(\imath_{u} F) ~~, \nn\\
  \delta S_{\mu}
  &= \mathfrak{D} \mu T h_{\{\mu\}}~~,
\end{align}
parametrised by $q$+8 transport coefficients for $q>1$ and $8$ for $q=1$,
namely, 2 shear viscosities $\eta_\Pi$, $\eta_\Gamma$, 4 bulk viscosities
$\zeta_\Pi$, $\zeta_\Gamma$, $\zeta_{\Pi,\Gamma}$, $\zeta_{\Gamma,\Pi}$, 1
diffusion constant $\mathfrak{D}$ and $q+1$ kinds of resistivities $\Theta$,
$\Omega_{I}$ satisfying
\begin{equation}
  \left(\eta_\Pi,\eta_\Gamma,\mathfrak{D},\Theta,
    \Omega_{I},\zeta_{\Gamma},\zeta_\Pi\right)\ge0~~,\qquad
  q(p-q)\zeta_{\Gamma}\zeta_\Pi\ge
  \frac{1}{4}\left((p-q)\zeta_{\Gamma,\Pi}+q\zeta_{\Pi,\Gamma}\right)^2~~.
\end{equation}
In the next section we will show that Onsager's relation implies that the bulk
cross-viscosities must be equal, i.e.,
$\zeta_{\Pi,\Gamma}=\zeta_{\Gamma,\Pi}=\zeta$ as also was shown to be the case
for $q=1$ in~\cite{Grozdanov:2016tdf, Hernandez:2017mch}.

It is worth noting that compared to the case $q=1$ studied
in~\cite{Grozdanov:2016tdf, Hernandez:2017mch}, there are $q$ additional
transport coefficients for $q>1$, namely the shear viscosity $\eta_\Pi$ in the
$\text{SO}(q)$ sector and $q-1$ resistivities $\Omega_{I\geq2}$. In the case
$q=1$ the shear viscosity in the $\text{SO}(q)$ sector is obsolete since
$\mathscr{L}_u \Pi_{\langle\mu\nu\rangle}=0$ as there is only one vector
$v^\mu_1$.


\subsection{Kubo formulae}

Here we derive Kubo formulae for the transport coefficients found above using
the variational background method of~\cite{Son:2007vk}. The results presented
here can be seen as an extension of those derived in~\cite{Grozdanov:2016tdf,
  Hernandez:2017mch} for $q=1$, though in a different frame. These formulae,
together with Onsager's relation imply a constraint between the two bulk
cross-viscosities found above, leading to one less independent transport coefficient.

We split the background coordinates $x^\alpha$ into the set $t$, $y^{a}$,
$z^{i}$ where $t$ is the time coordinate, $z^i$ with $i=1,\ldots,q$ label the
$q$ directions along $v_{i}^{\mu}$, and $y^{a}$ with $a=1,\ldots,p-q$ label the
orthogonal directions to $u^{\mu}$ and $v_{i}^{\mu}$. We consider an equilibrium
fluid configuration in flat space with vanishing background gauge field
$A_{q+1}$ and with velocity profile
\begin{equation}
  u^{\mu} = \delta^\mu{}_t~~,\qquad
  v^\mu_i = \delta^\mu{}_{i}~~.
\end{equation}
Performing a small time-dependent, but homogeneous in space, perturbation of all
fields $u^{\mu} \to u^{\mu} + \delta u^{\mu}$, $v_{i}^{\mu} \to v_{i}^{\mu}
+ \delta v_{i}{}^{\mu}$, $g_{\mu\nu} \to \eta_{\mu\nu} + \delta h_{\mu\nu}$ and $A_{q+1}\to\delta A_{q+1}$ leads to
\begin{equation} \label{eq:per}
  \delta u^t = \frac{1}{2}\delta h_{tt}~~,\qquad
  \delta v_i^t = \delta u^{i} + \delta h_{ti}~~,\qquad
  \delta_{k(j}\delta v_{i)}{}^k = -\frac{1}{2}\delta h_{ij}~~,
\end{equation}
as well as $\nabla_{(a}u_{b)}=\partial_t \delta h_{ab}/2$ and
$\nabla_{(i}u_{j)} = \partial_t \delta h_{ij}/2$. We define the one-point
functions
\begin{equation}
  \mathbb{T}^{\mu\nu}
  = \sqrt{-g}~\langle T^{\mu\nu}\rangle~~,\qquad
  \mathbb{J}^{\mu_1\ldots\mu_{q+1}}
  = \sqrt{-g}~\langle J^{\mu_1\ldots\mu_{q+1}}\rangle~~,
\end{equation}
which according to linear response theory can be written in terms of retarded
Green's functions of frequency $\omega$ such that (see
e.g.~\cite{Kovtun:2012rj})
\begin{equation}
\begin{split}
  \delta \mathbb{T}^{\mu\nu}(\omega)
  &= \frac{1}{2}G_{TT}^{\mu\nu,\lambda\rho}(\omega) \delta h_{\lambda\rho}
  + \frac{1}{(q+1)!} G_{TJ}^{\mu\nu,\mu_1\ldots\mu_{q+1}}(\omega)
  \delta (A_{q+1})_{\mu_1\ldots\mu_{q+1}}~,\\
  \delta \mathbb{J}^{\mu_1\ldots\mu_{q+1}}(\omega)
  &= \frac{1}{2}G_{JT}^{\mu_1\ldots\mu_{q+1},\lambda\rho}\delta h_{\lambda\rho}(\omega)
  + \frac{1}{(q+1)!} G_{JJ}^{\mu_1\ldots\mu_{q+1},\nu_{1}\ldots\nu_{q+1}}(\omega)\delta
  (A_{q+1})_{\nu_1\ldots\nu_{q+1}}~.
\end{split}
\end{equation}
Using~\eqref{eq:per} and the form of the stress tensor and charge current~\eqref{eq:1storder} one finds
\begin{equation}
\begin{split}
  \eta_{\Gamma}
  & = \lim_{\omega\to0}\frac1\omega ~\mathrm{Im}~ G_{TT}^{ab,ab}~,~
  \text{for}~a\ne b~~,\qquad
  \zeta_\Gamma+2\frac{(p-q-1)}{(p-q)}\eta_\Gamma
  = \lim_{\omega\to0}\frac1\omega ~\mathrm{Im}~G_{TT}^{aa,aa}~~, \\
  \eta_\Pi
  &= \lim_{\omega\to0}\frac1\omega ~\mathrm{Im}~ G_{TT}^{ij,ij}~,~
  \text{for}~i\ne j~~, \qquad
  \zeta_\Pi + \frac{2(q-1)}{q}\eta_\Pi
  = \lim_{\omega\to0}\frac1\omega ~\mathrm{Im}~G_{TT}^{ii,ii}~~, \\
  \zeta
  &= \zeta_{\Pi,\Gamma} = \zeta_{\Gamma,\Pi}
  = \lim_{\omega\to0}\frac1\omega ~\mathrm{Im}~G_{TT}^{ii,aa}
  = \lim_{\omega\to0}\frac1\omega ~\mathrm{Im}~G_{TT}^{aa,ii}~~, \\
  \mathfrak{D}
  &= \lim_{\omega\to0}\frac1\omega ~\mathrm{Im}~
  G_{JJ}^{ai_{1}\ldots i_{q},ai_{1}\ldots i_{q}}~~, \nn\\
  \Omega_{I}
  &= \lim_{\omega\to0}\frac1\omega ~\mathrm{Im}~
  G_{JJ}^{a_{1}\ldots a_{I+1} i_1\ldots i_{q-I}, a_{1}\ldots a_{I+1} i_1\ldots i_{q-I}}~~, \\
  \Theta
  &= \lim_{\omega\to0}\frac1\omega ~\mathrm{Im}~
  G_{JT}^{ati_{1}\ldots i_{q-1},ai}~~.
\end{split}
\end{equation}
Note that in the third line we have identified
$\zeta=\zeta_{\Pi,\Gamma}=\zeta_{\Gamma,\Pi}$. This follows from Onsager's
relation for mixed correlation functions, exactly in the same manner as
in~\cite{Grozdanov:2016tdf, Hernandez:2017mch} for $q=1$. This implies that
parity-even fluids with $q$-form symmetry are characterised by $q$+7 independent
transport coefficients at first order in derivatives for $q>1$ and 7 independent
transport coefficients for $q=1$.


\subsection{Constraints on transport in the isotropic limits}

As stressed in Sec.~\ref{sec:ideal}, the hydrodynamics of fluids with $q$-form symmetry must reduce to the hydrodynamics of (charged) isotropic fluids in the limits $q=0$, $q=p$ or $Q\to0$. This requirement imposes relations between transport coefficients in the isotropic limit. The corrections to the stress tensor and charge current~\eqref{eq:1storder} in the isotropic limit must take the following form in the Landau frame (see e.g.~\cite{Gath:2013qya, DiDato:2015dia}) 
\begin{equation} \label{eq:stiso}
\begin{split}
&\delta T^{\mu\nu}=-\eta{\Delta_\mu}^{\lambda}{\Delta_\nu}^{\rho}\left(\mathscr{L}_u g_{\lambda\rho}-\frac{1}{p}g_{\lambda\rho}\text{Tr}\mathscr{L}_u g\right)-\xi \vartheta_p \Delta^{\mu\nu}~~,\\
&\delta J^\mu=-\mathcal{D}T\Delta^{\mu\nu}\left(\nabla_\nu\left(\frac{\mu}{T}\right)-\frac{1}{T}E_\nu\right)~~,
\end{split}
\end{equation}
for some coefficients $\eta$, $\xi$, $\mathcal{D}$ and where we have only
considered corrections to the current for the $q=0$ case since the total current
vanishes in the limit $Q\to0$ while the corrections $\delta J$ vanish when $q=p$
since in this case $Q$ is a global charge and hence non-dynamical.

The most non-trivial limit is the uncharged (neutral) limit $Q\to0$ which must remove all sources of anisotropy. This means that when rewriting the stress tensor in~\eqref{eq:1storder} in terms of a $\text{SO}(p)$ and a $\text{SO}(q)$ sector, the $\text{SO}(q)$ sector must vanish in the limit $Q\to0$. This imposes non-trivial constraints on the transport coefficients in this limit. In particular we must have
\begin{equation} \label{eq:etazeta}
  \eta^{(0)}_\Pi=\eta^{(0)}_\Gamma~~,\qquad
  \zeta^{(0)}=\frac{(p-q)\zeta^{(0)}_\Gamma-q\zeta^{(0)}_\Pi}{p-2q}~~,
\end{equation}
where the subscript $(0)$ denotes the value of the transport coefficients in the limit $Q\to0$. In addition to~\eqref{eq:etazeta} one also obtains
\begin{equation} \label{eq:etazeta1}
  \zeta^{(0)}_\Gamma-\zeta^{(0)}-2\frac{\eta_\Gamma^{(0)}}{(p-q)}=0~~,\qquad
  \zeta^{(0)}-(\zeta^{(0)}_\Pi + \frac{1}{2}\zeta^{(0)}_\Gamma)
  + \frac{p}{q(p-q)}\eta_\Pi^{(0)}=0~~.
\end{equation}
However, the second condition above is redundant as the first condition in~\eqref{eq:etazeta1} together with conditions~\eqref{eq:etazeta} imply the second in~\eqref{eq:etazeta1}. Given these relations, comparison with~\eqref{eq:stiso} yields 
\begin{equation} \label{eq:t01}
  \eta=\eta^{(0)}~~,\qquad
  \xi=\zeta_\Gamma^{(0)}-\frac{q}{p(p-q)}\eta_\Gamma^{(0)}~~.
\end{equation}
As we shall see in a companion paper where we study a specific class of
gravitational duals to these fluid configurations in flat
space~\cite{toappear1}, the first condition in~\eqref{eq:etazeta} is actually
satisfied for any value of $p$, $q$, $Q$ since it is found that
$\eta_\Pi/S=\eta_\Gamma/S=1/4\pi$. The second condition in~\eqref{eq:etazeta} is
also satisfied for all $p$, $q$, $Q$ for the same class of gravitational
duals. This latter relation is also observed in the context of another class of
gravitational duals in Anti-de Sitter space studied in~\cite{Grozdanov:2017kyl},
for which it was found $\zeta_\Gamma=\zeta_\Pi/4=-\zeta/2$ (see Eq.\ (88)
in~\cite{Grozdanov:2017kyl}) in agreement with the second condition
in~\eqref{eq:etazeta} for $(p,q)=(3,1)$.

In the limit $q=0$ for which the fluid is carrying particle charge, we obtain the identification
\begin{equation} \label{eq:t11}
  \eta=\eta_{\Gamma}^{(q=0)}~~,\qquad
  \xi=\zeta_\Gamma^{(q=0)}~~,\qquad
  \mathcal{D}=\mathfrak{D}^{(q=0)}~~,
\end{equation}
while in the limit $q=p$ we find
\begin{equation} \label{eq:t21}
  \eta=\eta_{\Pi}^{(q=p)}~~,\qquad
  \xi=\zeta_\Pi^{(q=p)}~~.
\end{equation}
Taking the subsequent limit $Q\to0$ of the viscosities $\eta$ and $\xi$ in~\eqref{eq:t11}-\eqref{eq:t21} must lead to the results~\eqref{eq:t01} so that
\begin{equation}
  \eta=\eta_{\Pi}^{(q=p)}|_{Q\to0}=\eta_{\Gamma}^{(q=0)}|_{Q\to0}=\eta^{(0)}~~,\quad
  \xi=\zeta_\Pi^{(q=p)}|_{Q\to0} = \zeta_\Gamma^{(q=0)}|_{Q\to0}
  = \zeta_\Gamma^{(0)}-\frac{q}{p(p-q)}\eta_\Gamma^{(0)}~~.
\end{equation}
As will be shown in a companion publication~\cite{toappear1}, it is possible to
verify these relations using a family of gravitational duals parametrised by
$p$, $q$ and $Q$.


\section{Equilibrium partition function}\label{sec:equilibrium}

In this section we study the hydrostatic sector of the hydrodynamic theory
presented in the previous sections. We show that the most general partition
function that gives rise to the fluids with $q$-form symmetry introduced above,
regardless of the microscopic origin of the vectors $v_i^{\mu}$, requires the
existence of $q$ mutually commuting spacelike isometries in addition to a
timelike isometry. This analysis shows that, under these assumptions,
the most general partition function is a $q$-form generalisation of the free
energy for $q=1$ fluids considered in~\cite{Caldarelli:2010xz}. At the end of
this section, we show that further constraints must be imposed on the partition
function in order to describe the hydrostatic sector of the theory.


\subsection{The partition function}

In order to describe equilibrium solutions, we consider the existence of a
timelike Killing vector field $\textbf{k}^\mu$ with modulus
$\textbf{k}=|{-}g_{\mu\nu}\textbf{k}^\mu\textbf{k}^\nu|^{1/2}$. For the
configuration to be stationary, the Lie derivative along $\textbf{k}^\mu$ of any
quantity characterising the fluid must vanish, in particular
\begin{equation}
  \mathscr{L}_{\textbf{k}} g_{\mu\nu}=0~~,\qquad
  \mathscr{L}_{\textbf{k}} A_{q+1}=0~~.
\end{equation}
In order to construct the partition function we must classify the ideal order
Lorentz and gauge invariants on which the partition may depend on. It is
straightforward to realise that there are only two possible scalars that can be
constructed from background data, namely
\begin{equation}\label{q0scalars}
  \textbf{k}^2~~,\qquad
  \textbf{k}^{\mu}A_\mu\quad\text{for}\quad q=0~~,
\end{equation}
where the second scalar is only defined for $q=0$. Note that under a time
independent gauge transformation $\Lambda$ with
$\mathbf k^{\mu}\dow_{\mu}\Lambda = 0$, the second scalar is indeed
gauge-invariant. Requiring the partition function to be dependent on these two
scalars leads to the stress tensor, currents and thermodynamics of a $q=0$
charged fluid.

For $q>0$, however, there is no natural equivalent of the second scalar in
\eqref{q0scalars} since the respective contraction
$\mathbf k^{\mu}A^{q+1}_{\mu\mu_{1}\mu_{2}\ldots\mu_{q}}$ is a $q$-form.
Consequently, to describe the hydrostatic sector of fluids with a $q$-form
symmetry, we need to introduce additional tensor structures on the
background. If we were provided with a $(q+1)$-dimensional Killing
subspace, we could use the associated volume form to contract with the indices
of $A^{q+1}_{\mu_{1}\mu_{2}\ldots\mu_{q+1}}$ and obtain a gauge-invariant
scalar. To this end, we assume the existence of $q$ mutually commuting spacelike
Killing vectors $\boldsymbol{\ell}^\mu_i$ that satisfy
$[\textbf{k}^\mu,\boldsymbol{\ell}^\nu_i]=0$, but whose inner product
$\textbf{k}^\mu \boldsymbol{\ell}^\nu_i g_{\mu\nu}$ does not necessarily
vanish. Since the vectors $\boldsymbol{\ell}^\mu_i$ are Killing vectors fields, one must
also have that
\begin{equation}
  \mathscr{L}_{\boldsymbol{\ell}_i} g_{\mu\nu}=0~~,\qquad
  \mathscr{L}_{\boldsymbol{\ell}_i} A_{q+1}=0~~,\qquad
  \mathscr{L}_{\boldsymbol{\ell}_i} \textbf{k}^{\mu} = 0~~.
\end{equation}
The subspace spanned by the vectors $\bm\ell^{\mu}_{i}$ and $\mathbf k^{\mu}$ is the desired
$(q+1)$-dimensional Killing subspace, which is invariant under an arbitrary
redefinition of $\boldsymbol{\ell}_i^\mu$ involving themselves and
$\mathbf k^{\mu}$, i.e.
\begin{equation}\label{PF_extra_redef}
  \boldsymbol{\ell}_i^\mu\to \tensor{R}{_i^j}\boldsymbol{\ell}_j^\mu
  + P_i \textbf{k}^\mu~~.
\end{equation}
Here $\tensor{R}{_i^j}$ is an arbitrary $q\times q$ non-singular matrix and
$P_i$ is a $q$-vector. In addition to the timelike Killing vector field
$\mathbf k^{\mu}$, we require the equilibrium partition function to only
depend on the $(q+1)$-dimensional Killing subspace, and not on the Killing
vector fields $\bm\ell^{\mu}_{i}$ individually. This is equivalent to demanding the invariance of the partition function
under \eqref{PF_extra_redef}. As it will be shown below, this requirement leads to the correct 
constitutive relations for $q$-form hydrodynamics as introduced in Sec.~\ref{sec:ideal}.


It is convenient to fix a large part of the redefinition freedom
\eqref{PF_extra_redef} by instead working with a set of orthonormal vectors
defined as
\begin{equation}
  \zeta^{\mu}_i = \tensor{S}{_{i}^{j}}
  \lb \boldsymbol{\ell}_j^{\mu}-\frac{\textbf{k}_\nu
    \boldsymbol{\ell}_j^\nu }{\textbf{k}^2} \textbf{k}^\mu\rb~~,\qquad
  \text{where} \quad
  \tensor{S}{_{i}^{k}} \tensor{S}{_{j}^{l}}
  \boldsymbol{\ell}_k^{\mu}\boldsymbol{\ell}_l^{\nu}
  \lb  g_{\mu\nu} - \frac{\textbf{k}_\mu\textbf{k}_\nu}{\textbf{k}^2} \rb
  = \delta_{ij}~~,
\end{equation}
which satisfy $\zeta^{\mu}_i \textbf{k}_\mu=0$ and
$\zeta^{\mu}_{i}\zeta^{\nu}_{j}g_{\mu\nu} = \delta_{ij}$, and span a
$q$-dimensional subspace transverse to $\textbf{k}^{\mu}$. The second condition
above can be seen as determining the matrix $\tensor{S}{_{i}^{j}}$ up to a
residual $\SO(q)$ symmetry which rotates the vectors $\zeta^{\mu}_{i}$. This is
precisely the $\SO(q)$ symmetry of $q$-form hydrodynamics as introduced in
Sec.~\ref{sec:ideal} (see footnote~\ref{symmetry_footnote}).  Imposing the
normalisation conditions removes $q + q(q+1)/2$ components from $\zeta^{\mu}_i$,
while the $\SO(q)$ symmetry removes further $q(q-1)/2$ components. This leads to
$(p-q)q$ independent components in $\zeta^{\mu}_i$, matching the counting
performed in Sec.~\ref{sec:ideal} for the independent components of the fields
$v^{\mu}_{i}$. As a trade-off for working in an orthonormal basis, the
$\zeta^{\mu}_i$'s are not necessarily Killing vector fields, since while
$\boldsymbol{\ell}_i^\nu \textbf{k}_\nu/\textbf{k}^2$ cannot depend on the
directions along the timelike and spacelike isometries, it may depend on the
transverse $(p-q)$ directions.

Given these considerations, it follows that there are now two scalars that can
be built from background data and on which the partition function may depend on,
namely\footnote{The scalar $\sigma$ is the pullback of the background gauge
  field $A_{q+1}$ onto the $(q+1)$-dimensional subspace spanned by the vectors
  $\textbf{k}^\mu$ and $\boldsymbol{\ell}_i^\mu$. This scalar is gauge invariant
  due to the Killing properties of $\textbf{k}^\mu$ and
  $\boldsymbol{\ell}_i^\mu$. On the other hand, zero-derivative scalars
  constructed from the projection of $A_{q+1}$ onto the $(p-q)$-dimensional
  subspace will not be gauge invariant for arbitrary $p$ and $q$.}
\begin{equation}
  \textbf{k}^2~~,~~
  \sigma\equiv \frac{1}{q!}\epsilon^{j
    i_1\ldots i_{q-1}}\zeta^{\mu_1}_{i_1}\ldots \zeta^{\mu_{q-1}}_{i_{q-1}}
  \textbf{k}^\mu\zeta^\nu_{j}A^{q+1}_{\mu\nu\mu_1\ldots \mu_{q-1}}~~,
\end{equation}
which are well defined for all $p$ and $q$.\footnote{One may shift the scalar
  $\sigma$ by a constant $\mu_0$ which would turn out to have the interpretation
  of a constant global chemical potential.} In defining $\sigma$, we have
introduced a $\SO(q)$ covariant Levi-Civita tensor with
$\epsilon_{123\ldots q} = 1$. We can now consider the partition function to be a
function of these two scalars, therefore we write\footnote{If we had not
  required the partition function to be invariant under~\eqref{PF_extra_redef},
  it could depend on many other scalars such as
  $f_{i} = \bm\ell^{\mu}_{i} \mathbf k_{\mu}$ and
  $h_{ij} = \bm\ell^{\mu}_{i}\bm\ell^{\nu}_{j} (g_{\mu\nu} - \mathbf
  k_{\mu}\mathbf k_{\nu}/\mathbf k^{2})$ for all $i,j=1,2,\ldots,q$, in addition
  to $\textbf{k}^2$ and $\sigma$. Moreover, it would have also been possible
  to consider $SO(q)$ invariant scalars that are not invariant under arbitrary redifinitions given by \eqref{PF_extra_redef}.
  This class of scalars includes $f_{i}f_{j} h^{ij}$ and $\det h_{ij}$ where $h^{ij}$ is the inverse of
  $h_{ij}$. All these possible extra dependences, though perhaps of interest for the description of other physical systems, would be incompatible with the $q$-form fluid 
  that we are trying to describe, whose thermodynamic properties only depend 
  on two scalar degrees of freedom, namely, $T$ and $\mu$.}
\begin{equation} \label{eq:partition}
  W=\frac{1}{T_0}\int_{\Sigma}d^{D-1}x\sqrt{-g}
  ~ \mathcal{P}\left(\textbf{k}^2,\sigma\right)~~,
\end{equation}
where $T_0$ is the constant global temperature and the integral is taken over a spatial hypersurface $\Sigma$. From this partition function we may obtain the stress tensor and currents in the following manner
\begin{equation} \label{eq:stpart}
  \begin{split} 
    T^{\mu\nu}
    &=\frac{2}{\sqrt{-g}}\frac {\delta W}{\delta g_{\mu\nu}}
    =\mathcal{P}g^{\mu\nu}
    + 2\frac{\partial \mathcal{P}}{\partial
      \textbf{k}^2}\textbf{k}^\mu\textbf{k}^\nu
    - \sigma \frac{\partial\mathcal{P}}{\partial\sigma}
    \delta^{ij}\zeta^{\mu}_i\zeta^{\nu}_j~~,\\
    J^{\mu_1\ldots \mu_{q+1}}&=\frac{1}{\sqrt{-g}}\frac {\delta W}{\delta A^{q+1}_{\mu_1\ldots \mu_{q+1}}}=(q+1)\frac{\partial\mathcal{P}}{\partial\sigma}\textbf{k}^{[\mu_1}\zeta^{\mu_2}_{i_1}\ldots \zeta^{\mu_{q+1]}}_{i_q}\epsilon^{i_1\ldots i_q}~~.
  \end{split}
\end{equation}
Comparing this stress tensor and current with~\eqref{eq:cur0}
and~\eqref{eq:stress0} one identifies
\begin{equation} \label{eq:id}
  P = \mathcal{P}~~,~~
  Q = \textbf{k}\frac{\partial\mathcal{P}}{\partial\sigma}~~,~~
  u^{\mu} = \frac{\textbf{k}^\mu}{\textbf{k}}~~,~~
  \mu = \frac{\sigma}{\textbf{k}}~~,~~
  E+P=2\textbf{k}^2\frac{\partial \mathcal{P}}{\partial \textbf{k}^2}~~,~~
  v^\mu_i = \zeta^{\nu}_i~~.
\end{equation}
In addition, by obtaining the total entropy from the partition function
\begin{equation}
S_{\text{tot}}=\frac{\partial\left(T_0W\right)}{\partial T_0}=\int_{\Sigma}d^{D-1}x\sqrt{-g}\frac{\partial \mathcal{P}}{\partial T_0}~~,
\end{equation}
and comparing with that obtained by integrating the ideal order entropy
current~\eqref{eq:ent0} leads to the identification $T=T_0/\textbf{k}$, which
together with~\eqref{eq:id}, yield the thermodynamic
properties~\eqref{eq:gd}. The solution defined by~\eqref{eq:id} and
$T=T_0/\textbf{k}$ agrees with that obtained in~\cite{Caldarelli:2010xz} for
$q=1$ which reduces to the one considered in~\cite{Grozdanov:2016tdf} only when
$\textbf{k}_\mu\boldsymbol{\ell}^\mu_1=0$.


\subsection{Additional constraints from the second law of thermodynamics}\label{newconstraints}

In equilibrium, a hydrodynamic theory is expected not to produce any entropy. If
the solution~\eqref{eq:id} provided by the equilibrium partition function above
is truly an equilibrium solution, then not only must the ideal order fluid
equations~\eqref{eq:current0},~\eqref{eq:commuting} and~\eqref{eq:stcon1} be
trivially satisfied, but also the divergence of the entropy current
\eqref{eq:divSframeless} must vanish so that no entropy is produced. We can
isolate all the independent tensor structures appearing in the entropy current
divergence and first order constitutive relations in terms of
\begin{equation}\label{offshellLike}
  \N_{(\mu}\bfrac{u_{\nu)}}{T}, \qquad
  "dd\lb\frac{\mu}{T}\text{Vol}_q\rb + \frac1T \imath_u F~~.
\end{equation}
All the other tensor structures appearing in Sec.~\ref{sec:diss} are given by
projections of these along $u^{\mu}$, $\Pi^{\mu\nu}$ and $\Gamma^{\mu\nu}$,
leading to
\begin{gather}
  u^{\mu}\nabla_{\mu}T~~, \quad
  \Gamma_{\mu}{}^{\nu}\lb \frac{1}{T}\nabla_{\nu}T + \mathfrak{a}_\nu\rb~~,\quad
  \Pi_{\mu}{}^{\nu}\lb \frac{1}{T}\nabla_{\nu}T + \mathfrak{a}_\nu\rb~~,\quad
  \vartheta_{(p-q)}~~,~~\vartheta_{q}~~, \nn\\
  \mathscr{L}_u \Gamma_{\{\mu\nu\}}~~,\quad
  \mathscr{L}_u \Pi_{\langle\mu\nu\rangle}~~,\quad
    \mathscr{L}_u \Gamma_{\{(\mu\nu)\rangle}
    + \mathscr{L}_u \Pi_{\{(\mu\nu)\rangle}~~, \nn\\
  u^{\mu}\nabla_{\mu}\bfrac{\mu}{T}~~,\quad
  h_{\{\mu\}}~~,\quad
  \text{dVol}_{q}^{U}~~,\quad
  \mu\,\text{dVol}^{\Gamma}_q + \Gamma^{2}(\imath_u F)~~, \quad
  \Gamma^{I+1}(\imath_{u}F)\big\vert_{I=2}^{q}~~.
  \label{non-hydrostatic-tensor-structures}
\end{gather}
One may readily check that the solution~\eqref{eq:id} together with
$T=T_0/\textbf{k}$ leads to the vanishing of the first tensor structure in
\eqref{offshellLike} but not the second. In components, this corresponds to the last two terms
in \eqref{non-hydrostatic-tensor-structures} , which do not vanish for the
solution~\eqref{eq:id}. From here it
follows that all the conservation laws are trivially satisfied, however, the
terms proportional to $\Omega_{I}$ in~\eqref{eq:1storder} contribute to entropy
production. To remedy this situation, we must require the second term in
\eqref{offshellLike} to vanish by hand. In equilibrium, this term evaluates
to\footnote{In the $q=1$ case, this condition reduces to
  \begin{equation} \nonumber
    \dow_{[\mu} \lb \frac{\mu}{T} v^{1}_{\nu]} \rb + \frac{u^{\lambda}}{T}
    F_{\lambda\mu\nu}
    = \frac{1}{T_{0}} \dow_{[\mu} \lb \sigma v^{1}_{\nu]} -
    \textbf{k}^{\rho}A_{|\rho|\nu]} \rb
    = \frac{1}{T_{0}} \dow_{[\mu} \lb
    \zeta^{1}_{\nu]}\zeta^{\sigma}_{1}
    \mathbf k^{\rho} A_{\rho\sigma}
    - \textbf{k}^{\rho}A_{|\rho|\nu]} \rb
    = 0~~.
  \end{equation}
  It is easy to see from here that this condition is generically non-trivial. To
  further supplement our intuition, let us choose a basis $(t,z,y^{a})$ such
  that $\mathbf k^{\mu} = \tensor{\delta}{^{\mu}_{t}}$ and
  $\bm\ell_{1}^{\mu} = \tensor{\delta}{^{\mu}_{z}}$. This renders the background
  metric and gauge field independent of the $t$ and $z$ coordinates. For simplicity,
  let us further choose $g_{\mu\nu} = \eta_{\mu\nu}$. The $[ab]$ components of
  the above equation then give a non-trivial condition
  \begin{equation} \nonumber
    \frac{1}{T_{0}} \dow_{[a} A_{b]t} = 0~~,
  \end{equation}
  which is clearly not satisfied for arbitrary $A_{\mu\nu}$. In the context of
  \cite{Grozdanov:2016tdf}, this leads to the vanishing of their eq. (3.12) in
  equilibrium, which would otherwise contribute to entropy production.  }
\begin{equation}\label{eq:back}
  "dd\lb\frac{\mu}{T}\text{Vol}_q\rb + \frac1T \imath_u F
  = \frac{1}{T_{0}} "dd\left(\sigma\mathrm{Vol}_q -
    \imath_{\textbf{k}} A_{q+1}\right) = 0~~,
\end{equation}
where we have used the identity that
$\mathscr{L}_{\textbf{k}} A_{q+1} = "dd (\imath_{\textbf{k}} A_{q+1}) +
\imath_{\textbf{k}}F$ vanishes in equilibrium. To have a consistent hydrostatic
solution, we must require this additional condition on our hydrostatic
backgrounds on top of the existence of $q$ additional spacelike
isometries.\footnote{Even though Ref.~\cite{Caldarelli:2010xz} did not
  impose~\eqref{eq:back} on their backgrounds, all their backgrounds do happen
  to satisfy~\eqref{eq:back}. This means that all dipole charged black hole
  configurations studied in~\cite{Caldarelli:2010xz} are indeed equilibrium
  solutions.}

To summarise, apart from the existence of a timelike isometry, we have introduced
two additional constraints on our backgrounds so that they admit a hydrostatic
solution: they must admit $q$ additional spacelike isometries and they must
satisfy the constraint~\eqref{eq:back}. This seems to be a feature of
hydrodynamics with higher-form symmetries. However, one may wonder if at higher
orders in the derivative expansion further conditions must be imposed on these
backgrounds to ensure the consistency of the hydrostatic sector so that no entropy is produced. 
Extending the all order analysis of the second law of thermodynamics given
in~\cite{Haehl:2015pja} to higher-form fluids, one can check that at arbitrarily
high derivative orders, the second law forces the entropy current divergence in
\eqref{eq:divSframeless} to be a positive semi-definite quadratic form made out
of various tensor structures in~\eqref{non-hydrostatic-tensor-structures} and
their derivatives. Since we have already ensured these to vanish on the
solution~\eqref{eq:id}, we are guaranteed to have a vanishing entropy current
divergence at arbitrarily high orders in the derivative expansion.

Let us use this opportunity to point out a rather unconventional feature of
hydrostatics in higher-form fluids as defined above. In 0-form hydrodynamics,
one generally finds that requiring the existence of an equilibrium partition
function does not give any new constraints on the constitutive relations,
besides those already imposed by the second law of thermodynamics
\cite{Bhattacharyya:2013lha, Bhattacharyya:2014bha}. In other words, requiring
the second law to hold on a set of constitutive relations is sufficient to
guarantee the existence of an equilibrium partition function. In the higher-form
case however, given that the existence of equilibrium relies on additional
spacelike isometries, it is worth investigating if this still holds in full
generality, in particular also when parity-invariance or charge conjugation
invariance are not imposed. For concreteness, let us focus on the case of fluids
with a 1-form symmetry and consider charge conjugation non-invariant derivative
corrections to the constitutive relations of the form
\begin{align}\label{problematicTerms}
  \delta T^{\mu\nu}
  &= \frac{\mu}{T^{2}} v_{1}^{\rho}\dow_{\rho}T
    \lb T \frac{\dow \alpha_1}{\dow T} u^{\mu}u^{\nu}
  - \frac{\mu}{T} \frac{\dow \alpha_1}{\dow (\mu/T)} v_{1}^{\mu}v_{1}^{\nu} \rb
    - 2 \alpha_1 \frac{\mu}{T^{2}} v_{1}^{(\mu} \N^{\nu)} T
  - \N_{\mu} \lb \alpha_1 \frac{\mu}{T} v_{1}^{\mu} \rb u^{\mu}u^{\nu}, \nn\\
  \delta J^{\mu\nu}
  &= \frac{2}{T} u^{[\mu} \lb \frac{\mu}{T^{2}} v_{1}^{\nu]} v_{1}^{\rho}\dow_{\rho}T
    \frac{\dow\alpha_1}{\dow \nu}
    + \alpha_1 \frac{1}{T} \N^{\nu]} T \rb, \nn\\
  \delta S^{\mu}
  &= 2 \alpha_1 ~\frac{\mu}{T^{3}} u^{[\rho} v_{1}^{\mu]}\dow_{\mu}T
    - \frac{1}{T} u_{\nu} \delta T^{\mu\nu}
    - \frac{\mu}{T} v^{1}_{\nu} \delta J^{\mu\nu},
\end{align}
where $\alpha_1(T,\mu)$ is some independent transport coefficient. These
expressions have been specifically engineered to satisfy the second law of
thermodynamics with no entropy production. Interestingly, on the supposed
solution~\eqref{eq:id} where $u^{\mu}$ and $v_{1}^{\mu}$ are aligned along
(linear combinations of) isometries, most of the terms vanish, but the second to
last term in the stress tensor in~\eqref{problematicTerms} and the last term in
the charge current remain. This is clearly in tension with the equilibrium
partition function because there are no first order parity-preserving (but
charge conjugation non-invariant) scalars that can be written in
equilibrium. Consequently, the partition function analysis sets $\alpha_1$ to
zero. It appears, therefore, that for fluids with a higher-form symmetry, the
equilibrium partition function analysis is imposing new constraints on the
transport coefficients, which do not follow from an entropy current
analysis. Although it is not a contradiction of any sort, it is in striking
contrast with 0-form hydrodynamics where, by itself, the requirement of the
second law to hold is sufficient to ensure a well-defined hydrostatic
sector.\footnote{The analysis of~\cite{Grozdanov:2016tdf} avoided these issues
  altogether by focusing on a sector which respects parity and
  charge-conjugation symmetry (which takes $v_{1}^{\mu}\to -v_{1}^{\mu}$). All
  the first order problematic terms of the kind~\eqref{problematicTerms} are not
  present if these requirements are imposed. However, it is not clear whether
  this continues to hold at higher orders in derivatives.} We will return to
these issues in a future publication~\cite{toappear2}.



\section{Surface dynamics of fluids with $q$-form symmetry}
\label{sec:surface}

In this section we study the surface transport properties of fluids carrying
$q$-brane charge following~\cite{Armas:2015ssd, Armas:2016xxg}. We first
introduce conservation equations for the surface dynamics and then generalise
the partition function discussed in the previous section in order to include the
presence of a surface. We then perform a surface entropy current analysis and
show that it agrees with the partition function expectations. Finally, we study
capillary waves on the surface of the fluid and find signatures of anisotropy in
the dispersion relation.

\subsection{Conservation equations and the second law of thermodynamics}

We introduce an interface/surface separating two different fluid phases by
adding localised source contributions to the stress tensor and
currents. Therefore, the full stress tensor, charge and entropy currents take
the form
\begin{equation} \label{eq:surcur}
\begin{split}
  T^{\mu\nu}
  &= T^{\mu\nu}_{\text{blk}}~\Theta(f)
  + T^{\mu\nu}_{\text{sur}}~\wt\delta(f)
  + \ldots ~~,\\
  J
  &= J_{\text{blk}}~\Theta(f)
  + J_{\text{sur}}~\wt\delta(f) + \ldots ~~,\\
  S^\mu
  &= S^\mu_{\text{blk}}~\Theta(f)
  + S^\mu_{\text{sur}}~\wt\delta(f) + \ldots ~~,
\end{split}
\end{equation}
where the \emph{dots} represent higher order corrections in the thickness
$\partial_\rho\delta(f)$ of the surface which we do not consider in the present
paper.\footnote{See~\cite{Armas:2015ssd} for a thorough analysis of these terms
  in the context of uncharged fluids.} In~\eqref{eq:surcur} we have introduced
the shape-field $f$ in terms of which the location of the surface is represented
as $f=0$. The step function $\Theta(f)$ vanishes at $f=0$ while
$\wt\delta(f)$ is the reparametrisation invariant delta function
$\wt\delta(f)=\sqrt{-\gamma}/\sqrt{-g} ~\delta(f)$. $\gamma$ denotes the
determinant of the induced metric on the surface,
$\gamma_{\mu\nu}=g_{\mu\nu}-n_\mu n_\nu$, and
$n_\mu=-\partial_\mu f/|\partial_\mu f\partial^\mu f|^{1/2}$ is the normal
co-vector to the surface.

The conservation equations for the stress tensor were already considered in~\cite{Armas:2015ssd, Armas:2016xxg} and the charge current conservation equation can be obtained by a simple generalisation of the results of~\cite{Armas:2012ac, Armas:2013aka}. These conservation laws take the form
\begin{equation} \label{eq:surlaws}
\wt\nabla_\mu {T^{\mu\nu}_{\text{sur}}}-F^{\mu}_{\text{sur}}=T^{\mu\nu}_{\text{blk}}n_\nu~~,~~\wt\nabla_\mu J^{\mu\mu_1\ldots \mu_q}_{\text{sur}}=J^{\mu\mu_1\ldots \mu_q}_{\text{bulk}}n_\mu~~,
\end{equation}
with
$F^{\mu}_{\text{sur}}={F^{\mu}}_{\mu_1\ldots \mu_{q+1}}J^{\mu_1\ldots
  \mu_{q+1}}_{\text{sur}}/(q+1)! $ and in addition the system must obey the
second law of thermodynamics
\begin{equation} \label{eq:2lawsur}
  \wt\nabla_\mu S^{\mu}_{\text{sur}}-S^{\mu}_{\text{blk}}n_\mu\ge0~~,
\end{equation}
together with the constraints
$T^{\mu\nu}_{\text{sur}}n_\mu = J^{\mu\mu_1\ldots \mu_q}_{\text{sur}}n_\mu =
S^{\mu}_{\text{sur}}n_\mu=0$. In Eq.~\eqref{eq:surlaws}, we have introduced the
surface projection of the background covariant derivative
$\wt\nabla_\mu\equiv{\gamma_\mu}^\nu\nabla_\nu$.\footnote{This covariant
  derivate should not be confused with the surface covariant derivative
  $\tilde\nabla_\mu$ introduced in~\cite{Armas:2016xxg}. The purpose of using
  the surface projection of the covariant derivative instead is to avoid having
  to work with the singular character of the delta function $\wt\delta(x)$.}
In the present paper, we choose a consistent truncation, as explained
in~\cite{Armas:2015ssd, Armas:2016xxg}, in which the bulk currents are expanded
up to first order in derivatives and the surface is kept at ideal order. This
implies that we take the bulk stress tensor and currents to be those derived in
the previous section at first order in derivatives, that
is~\eqref{eq:cur0},~\eqref{eq:stress0},~\eqref{eq:ent0} together
with~\eqref{eq:1storder}. We will now consider equilibrium configurations and
then perform a surface entropy current analysis.

\subsection{Equilibrium partition function and entropy current analysis}

Analogously to the cases studied in~\cite{Armas:2015ssd, Armas:2016xxg}, one may
write equilibrium partition functions for fluids with $q$-form symmetry in the
presence of surfaces. Up to first order in derivatives, and under the
assumptions of parity-invariance and charge conjugation invariance, the
partition function takes the form\footnote{Note that the
  condition~\eqref{eq:back} must be imposed on the background and a similar
  condition must be imposed on the surface in order for~\eqref{eq:part1} to be
  an equilibrium partition function.}
\begin{equation} \label{eq:part1} 
  W
  = \frac{1}{T_0}\int_{\Sigma}d^{D-1}x\sqrt{-g}~
  \mathcal{P}(\textbf{k}^2,\sigma)
  + \frac{1}{T_0}\int_{\wt\Sigma}d^{D-2}\wt x\sqrt{-\gamma}~
  \mathcal{C}(\textbf{k}^2,\sigma)~~,
\end{equation}
where $\Sigma$ is now a spatial hypersurface enclosed by the spatial
codimension-2 surface $\wt\Sigma$ at the boundary with coordinates $\wt
x$. Since we aim to describe stationary configurations, we must have
\begin{equation}
  \mathscr{L}_{\textbf{k}} f=0~~,~~\mathscr{L}_{\boldsymbol{\ell}_i} f=0~~.
\end{equation}
By performing a variation with respect to the background metric and gauge fields one obtains the stress tensor and currents in the form~\eqref{eq:surcur}, where, in particular, the surface stress tensor and charge current take the analogous form to~\eqref{eq:stpart} and with the exact same thermodynamic properties. We will explicitly derive these currents below using an entropy current analysis.

As shown in~\cite{Armas:2015ssd, Armas:2016xxg}, constraints on surface
transport can be obtained by analysing the divergence of the surface entropy
current. This analysis not only fixes the surface thermodynamics and currents,
but also the value of $u^\mu n_\mu$ at leading order on the surface. As stated
above, the surface currents must satisfy the constraints
$T^{\mu\nu}_{\text{sur}}n_\mu = J^{\mu\mu_1\ldots\mu_q}_{\text{sur}}n_\mu =
S^{\mu}_{\text{sur}}n_\mu=0$. This implies that at ideal order, and ignoring
parity-odd effects, the surface currents take the form\footnote{It is possible
  to consider a component in the surface stress tensor of the form
  $\wt{u}^{(\mu}\wt{v}^{\mu)}_i$ which would ultimately be required to vanish by
  the second law of thermodynamics. For clarity of presentation, we have not
  considered it.}
\begin{equation}
  \begin{split}
    T^{\mu\nu}_{\text{sur}}
    &= \left({\mathcal{E}}-{\mathcal{Y}}\right)\wt u^\mu \wt u^\nu
    - {\mathcal{Y}}~\wt{\Gamma}^{\mu\nu}
    - \lb {\mathcal{Y}} + \mu\mathcal{Q} \rb~\wt{\Pi}^{\mu\nu}~~,\\
    J_{\text{sur}}
    &= {\mathcal{Q}}~\wt{\text{Vol}}_{(q+1)}~~,\\
    S_{\text{sur}}^{\mu}
    &= {\mathcal{S}}~ \wt u^{\mu}~~.
  \end{split}
\end{equation}
Here $\wt u^\mu = u^\mu-(u.n)n^\mu$ and $\wt\Pi^{\mu\nu}$ is a projector
analogous to $\Pi^{\mu\nu}$ but constructed out the vectors
$\wt v_i^\mu = v_i^\mu-(v_i.n)n^\mu$. Similarly, $\wt{\text{Vol}}_{(q+1)}$ is
the volume form introduced in~\eqref{eq:cur0} but with $u^{\mu}$, $v^{\mu}_i$
replaced by $\wt u^{\mu}$, $\wt v^{\mu}_i$. These two vectors satisfy
$\wt u.n = \wt v_i.n=0$. The projector $\wt{\Gamma}^{\mu\nu}$ is
constructed using the induced metric so that
$\wt{\Gamma}^{\mu\nu} = \gamma^{\mu\nu} + \wt u^\mu \wt u^\nu
-\wt {\Pi}^{\mu\nu}$.

Noting that the bulk entropy current $S_{\text{blk}}^{\mu}$ is given
by \eqref{eq:ent0} and \eqref{eq:1storder}, requiring the second law~\eqref{eq:2lawsur} to be
satisfied leads to
\begin{equation}
  \frac{\wt{u}^\mu}{T} \left(T\nabla_\mu \mathcal{S}
    + \mu\nabla_\mu \mathcal{Q} - \nabla_\mu \mathcal{E}\right)
  - \frac{1}{T}\left(\mathcal{E} - \mathcal{Y} - T\mathcal{S} -
    \mu\mathcal{Q}\right)
  \gamma^{\mu\nu}\wt\nabla_\mu\wt u_\nu
  - \frac{u.n}{T} \delta T^{\mu\nu}_{\text{blk}}n_\mu n_\nu\ge0~~,
\end{equation}
where $\delta T^{\mu\nu}_{\text{blk}}$ is the first order correction to the bulk
stress tensor given in~\eqref{eq:1storder}. In the present case, requiring
positivity of the entropy current allows us to deduce the surface thermodynamics
\begin{equation}
  "dd \mathcal{E} = T "dd\mathcal{S} + \mu "dd\mathcal{Q}~~,\qquad
  \mathcal{E} - \mathcal{Y} = T\mathcal{S} + \mu\mathcal{Q}~~,
\end{equation}
as well as the expected condition $u.n=0$. This condition is expected to be
modified at higher-orders~\cite{Armas:2016xxg}. In particular, do note that this
analysis, similar to the case of the superfluid velocity
in~\cite{Armas:2016xxg}, does not fix the components $v_i.n$ at the surface. The
stress tensor, current and thermodynamics obtained here agree with those that
are readily derived from~\eqref{eq:part1}, once we identify $\mathcal{C}=-\mathcal{Y}$.


\subsection{Surface conservation equations}

Having derived the surface constitutive relations at ideal order, we can write down the first order equations of motion 
at the surface using \eqref{eq:surlaws}. After imposing $u\cdot n = 0$, the components of the $q$-form charge conservation law in
\eqref{eq:surlaws} along with  $\wt u^{\mu}$ and $\wt\Pi^{\mu\nu}$ are given by
\begin{equation}
  \begin{split}
    u^{\mu}\nabla_{\mu}\mathcal{Q} +
    \mathcal{Q}\gamma^{\mu\nu}\wt\nabla_{\mu}u_{\nu}
    - \mathcal{Q}\wt{\Pi}^{\mu\nu}\nabla_{\mu}u_{\nu} &= 0~~, \\
    \wt{\Pi}_{\alpha}{}^{\nu}\nabla_{\nu}\mathcal{Q} - \mathcal{Q}
    \wt{\Pi}_{\alpha\nu}u^{\mu}\nabla_{\mu}u^{\nu} + \mathcal{Q}
    \wt{\Pi}_{\alpha\nu}
    \wt\nabla_{\sigma}\wt{\Pi}^{\sigma\nu} &= Q
    \wt{v}^{i}_{\alpha}\lb v^{\mu}_{i}n_{\mu} \rb~~,
  \end{split}
\end{equation}
while those along $\wt{\Gamma}^{\mu\nu}$ are
\begin{equation} \label{eq:vanLie}
  \wt{\Gamma}_{\alpha\mu}\lb u^{\nu}\nabla_{\nu}\wt{v}_{i}^{\mu}
  - \wt{v}^{\nu}_{i}\nabla_{\nu}u^{\mu} \rb
  = \wt{\Gamma}_{\alpha\mu}\lb \wt{v}_{i}^{\nu}\nabla_{\nu}\wt{v}_{j}^{\mu}
  - \wt{v}^{\nu}_{j}\nabla_{\nu}\wt{v}_{i}^{\mu} \rb = 0~~.
\end{equation}
These should be contrasted with the respective bulk $q$-form conservation laws
in \eqref{eq:current0} and \eqref{eq:commuting}. The component of charge conservation along
$n_{\mu}$ trivially vanishes.  Upon using the charge
conservation equations, the stress tensor conservation equation in
\eqref{eq:surlaws} can be projected along $\wt{u}^{\mu}$,
$\wt{\Pi}^{\mu\nu}$ and $\wt{\Gamma}^{\mu\nu}$ such that
\begin{equation} \label{eq:surfeq}
  \begin{split}
    T \lb u^{\mu}\nabla_{\mu}\mathcal{S}
    + \mathcal{S} \gamma^{\mu\nu} \wt\nabla_{\mu}u_{\nu} \rb &= 0~~, \\
    T \mathcal{S} ~\wt{\Pi}_{\alpha}{}^{\nu} \lb \frac1T \nabla_{\nu}T +
    \mathfrak{a}_\nu \rb
    &= 0~~, \\
    (\mathcal{E}-\mathcal{Y}) \wt{\Gamma}_{\alpha}{}^{\nu} \lb
    \frac{1}{T}\nabla_{\nu}T + \mathfrak{a}_\nu \rb + T
    \mathcal{Q}\wt{\Gamma}_{\alpha}{}^{\nu} \lb \nabla_{\nu}\bfrac{\mu}{T} -
    \frac{\mu}{T}
    \gamma^{\mu}{}_{\sigma}\nabla_{\mu}\wt{\Pi}^{\sigma}{}_{\nu} \rb
    &= \wt\Gamma_{\alpha\nu}F^{\nu}_{\text{sur}}~~, \\
  \end{split}
\end{equation}
which are analogous to the bulk equations given in \eqref{eq:stcon1}. Finally,
projecting the stress tensor conservation equation along $n_{\mu}$ we get the
Young-Laplace equation
\begin{gather}
  - (\mathcal{E}-\mathcal{Y})u^{\mu}u^{\nu}\nabla_{\mu}n_{\nu}
  + \mathcal{Y}\nabla_{\mu}n^{\mu}
  + \mu\mathcal{Q}\wt{\Pi}^{\mu\nu}\nabla_{\mu}n_{\nu}
  = P - \mu Q \Pi^{\mu\nu}n_{\mu}n_{\nu}~~,
\end{gather}
which provides an equation of motion for the shape-field. We will
solve these equations at the linear level in the next subsection.

\subsection{Surface waves}

In this subsection we study the nature of linearised fluctuations about an
equilibrium configuration.  For simplicity, we work on a flat background
with metric $\eta_{\mu\nu}$ and vanishing gauge field. We pick a basis
$x^{\mu} = \{t,z^{i},y^{a},r\}$ and work around the equilibrium solution given
by\footnote{For simplicity, we have assumed that $f$ has no dependence on the $z^{i}$ coordinates.}
\begin{equation}
  u^{\mu} = \delta_{t}{}^{\mu}~~, ~~
  v_{i}{}^{\mu} = \delta_{i}{}^{\mu}~~, ~~
    T = T_{0}~~, ~~
  \mu = \mu_{0}~~, ~~
  f = r~~.
\end{equation}
It follows that
$n_{\mu} = - \delta^{r}{}_{\mu}$.  This solution obviously satisfies the bulk equations
of motion. To solve the surface equations of motion, we must further require
$P(T_{0}) = 0$. Performing a small perturbation around this
solution and using that $u^{\mu}n_{\mu} = 0$ implies
\begin{equation}
  \delta u^{t} = 0~~, ~~
  \delta v_{i}{}^{t} = \delta u_{i}~~, ~~
  \delta v_{i}{}^{j} = - \delta v^{j}{}_{i}~~, ~~
  \delta u^{r} = - \dow_{t}\delta f~~.
\end{equation}
Furthermore, the vanishing Lie derivative conditions \eqref{eq:vanLie} imply
\begin{align}
  \mathrm{L}_{1}
  &\equiv \dow_{t}\delta v_{i}{}^{a} - \dow_{i}\delta u^{a} = 0~~, ~~
    \dow_{i}\delta v_{j}{}^{a} - \dow_{j}\delta v_{i}{}^{a} = 0~~, \\
  \mathrm{L}_{2}
  &\equiv \dow_{t}\delta v_{i}{}^{r} - \dow_{i}\delta u^{r}
    = \dow_{t} \lb \delta v_{i}{}^{r} + \dow_{i}\delta f\rb = 0~~, ~~
    \dow_{i}\delta v_{j}{}^{r} - \dow_{j}\delta v_{i}{}^{r} = 0~~,
\end{align}
which can be seen as determining $\delta v_{i}{}^{a}$ and $\delta v_{i}{}^{r}$
respectively. Note that the antisymmetric modes in $\delta v_{i}{}^{j}$ are not
physical due to the underlying $\SO(q)$ symmetry. Therefore, the remaining degrees of freedom we need to
solve for are $\delta T$, $\delta\mu$, $\delta u^{i}$, $\delta u^{a}$
and $\delta f$. Let us first look at the boundary equations of motion. The
Young-Laplace condition provides an equation of motion for $\delta f$ that takes the form
\begin{equation}
  \mathrm{YL} \equiv \mathcal{E} \dow_{t}^{2} \delta f
  - \lb \mathcal{Y} + \mu\mathcal{Q} \rb \delta^{ij}\dow_{i}\dow_{j}\delta f
  - \mathcal{Y} \delta^{ab}\dow_{a}\dow_{b} \delta f
  - S \delta T - Q \delta \mu~~.
\end{equation}
The surface conservation equations~\eqref{eq:surfeq} imply
\begin{align}
  \mathrm{S}_{1} &\equiv \dow_{t}\delta \mathcal{S}
  + \mathcal{S} \dow_{i} \delta u^{i}
  + \mathcal{S} \dow_{a} \delta u^{a} = 0~~, \\
  \mathrm{S}_{2} &\equiv \frac1T \dow_{i} \delta T + \dow_{t} \delta u_{i}
  = 0~~, \\
  \mathrm{S}_{3} &\equiv (\mathcal{E}-\mathcal{Y})
  \lb \frac{1}{T}\dow_{a}\delta T + \dow_{t}\delta u_{a} \rb 
  + T \mathcal{Q} \lb \dow_{a}\delta \bfrac{\mu}{T}
  - \frac{\mu}{T}\dow_{i}\delta v^{i}{}_{a} \rb = 0~~, \\
  \mathrm{S}_{4} &\equiv \dow_{t} \delta \mathcal{Q}
  + \mathcal{Q}\dow_{a}\delta u^{a} = 0~~, \\
  \mathrm{S}_{5} &\equiv \dow_{i} \delta \mathcal{Q}
  + \mathcal{Q} \dow_{a}\delta v_{i}{}^{a}
  + Q \lb \delta v_{i}{}^{r} + \dow_{i}\delta f \rb = 0~~.
\end{align}
Finally, the bulk equations of motion \eqref{eq:current0},\eqref{eq:commuting} and \eqref{eq:stcon} at the linear level are given by
\begin{align}
  \mathrm{B}_{1}
  &\equiv \dow_{t} \delta S
    + S \dow_{i}\delta u^{i}
    + S \dow_{a}\delta u^{a}
    - S \dow_{t}\dow_{r}\delta f = 0~~, \\
  \mathrm{B}_{2}
  &\equiv \frac1T \dow_{i}\delta T + \dow_{t}\delta u_{i} = 0~~, \\
  \mathrm{B}_{3} &\equiv E
  \lb \frac{1}{T}\dow_{a}\delta T + \dow_{t}\delta u_{a} \rb 
  + TQ \lb \dow_{a}\delta \bfrac{\mu}{T}
  - \frac{\mu}{T} \dow_{i}\delta v^{i}{}_{a} \rb = 0~~, \\
  \mathrm{B}'_{3} &\equiv E
  \lb \frac{1}{T}\dow_{r}\delta T - \dow^{2}_{t}\delta f \rb 
  + TQ \lb \dow_{r}\delta \bfrac{\mu}{T}
  - \frac{\mu}{T} \dow_{i}\delta v^{i}{}_{r} \rb = 0~~, \\
  \mathrm{B}_{4}
  &\equiv \dow_{t}\delta Q + Q \dow_{a}\delta u^{a} - Q \dow_{t}\dow_{r}\delta f = 0~~, \\
  \mathrm{B}_{5}
  &\equiv \dow_{i} \delta Q
    + Q \dow_{a} \delta v_{i}{}^{a}
    + Q \dow_{r} \delta v_{i}{}^{r}
  = 0~~.
\end{align}
We focus on plane wave solutions to these equations which behave as
$\E{i(\omega t - k_{i}z^{i} - \ell_{a}y^{a})}\E{ -\kappa r}$, where $\omega$ is the frequency of the wave, $k_i$ is the wavenumber along the $(p-q)$ directions, $\ell_a$ the wavenumber along the $q$ anisotropic directions and $\kappa$ is a damping factor. Equations
$\mathrm{L}_{1}$, $\mathrm{L}_{2}$, $\mathrm{B}_{2}$, $\mathrm{S}_{2}$ are then immediately solved by choosing
\begin{equation}
  \delta v_{i}{}^{a} = - \frac{k_{i}}{\omega} \delta u^{a}~~, ~~
  \delta v_{i}{}^{r} = i k_{i} \delta f~~, ~~
  \delta u^{i} = \frac{k^{i}}{\omega} \frac{1}{T} \delta T~~.
\end{equation}
This also makes $\mathrm{B}_{5}$ and $\mathrm{S}_{5}$ linearly dependent on
$\mathrm{B}_{4}$ and $\mathrm{S}_{4}$ respectively. The $\delta u^{a}$ components
of the velocity are obtained by solving $\mathrm{B_{3}}$ and $\mathrm{S_{3}}$ such that
\begin{equation}
  \delta u^{a}
  = \frac{\omega \ell^{a}}{\omega^{2} E - k^{2} Q \mu}
  \lb S\delta T + Q \delta\mu \rb~~, ~~
  \delta u^{a} \big\vert_{r\to0}
  = \frac{\omega \ell^{a}}{\omega^{2} (\mathcal{E}-\mathcal{Y})
    - k^{2} \mathcal{Q} \mu}
  \lb \mathcal{S}\delta T + \mathcal{Q} \delta\mu \rb~~.
\end{equation}
Finally, we can turn to the scalar degrees of freedom $\delta T$ and $\delta \mu$ which are given by solutions of
$\mathrm{B}_{1}$, $\mathrm{B}'_{3}$, $\mathrm{B}_{4}$ in the bulk and
$\mathrm{S}_{1}$, $\mathrm{S}_{4}$ and $\mathrm{YL}$ at the surface. At the
surface we have
\begin{gather}
  \lb \frac{k^{2}}{\omega^{2}} \frac{\mathcal{S}}{T}
  + \frac{\ell^{2}\mathcal{S}^{2}}{\omega^{2} (\mathcal{E}-\mathcal{Y})
    - k^{2} \mathcal{Q} \mu}
  - \mathcal{X}_{TT} \rb \delta T
  + \lb \frac{\ell^{2}\mathcal{Q}\mathcal{S}}{\omega^{2} (\mathcal{E}-\mathcal{Y})
    - k^{2} \mathcal{Q} \mu}
  - \mathcal{X}_{T\mu} \rb\delta\mu = 0~~, \\
  \lb \frac{\ell^{2}\mathcal{Q}\mathcal{S}}{\omega^{2} (\mathcal{E}-\mathcal{Y})
    - k^{2} \mathcal{Q} \mu}
  - \mathcal{X}_{T\mu} \rb \delta T
  + \lb \frac{\ell^{2}\mathcal{Q}^{2}}{\omega^{2} (\mathcal{E}-\mathcal{Y})
    - k^{2} \mathcal{Q} \mu}
  - \mathcal{X}_{\mu\mu} \rb \delta\mu = 0~~, \\
  \Big( - \omega^{2} \mathcal{E}
  + k^{2} \mu\mathcal{Q}
  + \lb k^{2} + \ell^{2} \rb \mathcal{Y} \Big) \delta f
  = S \delta T + Q \delta \mu~~, \label{fSolBoundary}
\end{gather}
which provide boundary conditions for the bulk equations of motion
\begin{gather}
  \lb \frac{k^{2}}{\omega^{2}} \frac{S}{T}
  + \frac{(\ell^{2}-\kappa^{2})S^{2}}{\omega^{2} E - k^{2} Q \mu}
  - \chi_{TT}
  \rb \delta T
  + \lb \frac{(\ell^{2}-\kappa^{2})SQ}{\omega^{2} E - k^{2} Q \mu}
  - \chi_{T\mu}
  \rb \delta\mu = 0~~, \\
  \lb \frac{(\ell^{2}-\kappa^{2})QS}{\omega^{2} E - k^{2} Q \mu}
  - \chi_{T\mu} \rb \delta T
  + \lb \frac{(\ell^{2}-\kappa^{2})Q^{2}}{\omega^{2} E - k^{2} Q\mu}
  - \chi_{\mu\mu} \rb \delta\mu = 0~~, \\
  \delta f
  = \frac{\kappa}{\omega^{2}E - Q\mu k^{2}}
  \lb S \delta T + Q \delta \mu \rb~~. \label{fSolBulk}
\end{gather}
Here we have defined $k^2=k_i k^i$, $\ell^2=\ell_a\ell^a$ as well as the susceptibility matrices
\begin{gather}
  \mathcal{X}_{TT} = - \frac{\dow^{2}\mathcal{Y}}{\dow T^{2}} =
  \frac{\dow\mathcal{S}}{\dow T}~~, ~~
  \mathcal{X}_{T\mu} = - \frac{\dow^{2}\mathcal{Y}}{\dow T \dow\mu}
  = \frac{\dow\mathcal{S}}{\dow \mu}
  = \frac{\dow\mathcal{Q}}{\dow T}~~, ~~
  \mathcal{X}_{\mu\mu} = - \frac{\dow^{2}\mathcal{Y}}{\dow\mu^{2}} =
  \frac{\dow\mathcal{Q}}{\dow \mu}~~, \\
  \chi_{TT} = \frac{\dow^{2} P}{\dow T^{2}} = \frac{\dow S}{\dow T}~~, ~~
  \chi_{T\mu} = \frac{\dow^{2} P}{\dow T \dow\mu}
  = \frac{\dow S}{\dow \mu} = \frac{\dow Q}{\dow T}~~, ~~
  \chi_{\mu\mu} = \frac{\dow^{2} P}{\dow \mu^{2}} = \frac{\dow Q}{\dow \mu}~~.
\end{gather}

From Eqs.~\eqref{fSolBoundary} and \eqref{fSolBulk} we can read out the frequency $\omega$, which is given by
\begin{equation}\label{dispersion}
  \omega
  = \pm \sqrt{\frac{Q\mu k^{2} + \kappa k^{2} \mu\mathcal{Q}
  + \kappa \lb k^{2} + \ell^{2} \rb \mathcal{Y}}{E + \kappa\mathcal{E}}}~~,
\end{equation}
and is required for the consistency of the solution to $\delta f$. Finally, we have
consistency conditions involving $\delta T$ and $\delta\mu$, which will
determine $\kappa$, $k^{2}$ and $\ell^{2}$, namely
\begin{equation}\label{vanishingDet1}
  \begin{vmatrix}
    \frac{k^{2}}{\omega^{2}} \frac{S}{T}
    + \frac{(\ell^{2}-\kappa^{2})S^{2}}{\omega^{2} E - k^{2} Q \mu}
    - \chi_{TT} &
    \frac{(\ell^{2}-\kappa^{2})SQ}{\omega^{2} E - k^{2} Q \mu}
    - \chi_{T\mu} \\
    \frac{(\ell^{2}-\kappa^{2})SQ}{\omega^{2} E - k^{2} Q \mu}
    - \chi_{T\mu} &
    \frac{(\ell^{2}-\kappa^{2})Q^{2}}{\omega^{2} E - k^{2} Q\mu}
    - \chi_{\mu\mu}
  \end{vmatrix} = 0~~,
\end{equation}
and the following vanishing determinants at the surface
\begin{equation}\label{vanishingDet2}
  \begin{vmatrix}
    \frac{k^{2}}{\omega^{2}} \frac{\mathcal{S}}{T} +
    \frac{\ell^{2}\mathcal{S}^{2}}{\omega^{2} (\mathcal{E}-\mathcal{Y}) - k^{2}
      \mathcal{Q} \mu} - \mathcal{X}_{TT} &
    \frac{\ell^{2}\mathcal{Q}\mathcal{S}}{\omega^{2} (\mathcal{E}-\mathcal{Y}) -
      k^{2} \mathcal{Q} \mu}
    - \mathcal{X}_{T\mu} \\
    \frac{\ell^{2}\mathcal{Q}\mathcal{S}}{\omega^{2} (\mathcal{E}-\mathcal{Y}) -
      k^{2} \mathcal{Q} \mu} - \mathcal{X}_{T\mu} &
    \frac{\ell^{2}\mathcal{Q}^{2}}{\omega^{2} (\mathcal{E}-\mathcal{Y}) - k^{2}
      \mathcal{Q} \mu} - \mathcal{X}_{\mu\mu}
  \end{vmatrix} = 0~~,
\end{equation}
\begin{equation}\label{vanishingDet3}
  \begin{vmatrix}
    \frac{k^{2}}{\omega^{2}} \frac{S}{T}
    + \frac{(\ell^{2}-\kappa^{2})S^{2}}{\omega^{2} E - k^{2} Q \mu}
    - \chi_{TT} &
    \frac{(\ell^{2}-\kappa^{2})SQ}{\omega^{2} E - k^{2} Q \mu}
    - \chi_{T\mu} \\
    \frac{k^{2}}{\omega^{2}} \frac{\mathcal{S}}{T} +
    \frac{\ell^{2}\mathcal{S}^{2}}{\omega^{2} (\mathcal{E}-\mathcal{Y}) - k^{2}
      \mathcal{Q} \mu} - \mathcal{X}_{TT} &
    \frac{\ell^{2}\mathcal{Q}\mathcal{S}}{\omega^{2} (\mathcal{E}-\mathcal{Y}) -
      k^{2} \mathcal{Q} \mu}
    - \mathcal{X}_{T\mu}
  \end{vmatrix} = 0~~.
\end{equation}
\Cref{dispersion,vanishingDet1,vanishingDet2,vanishingDet3} completely
characterise the wave fluctuations of the $q$-form fluid with a surface.

We will now study the dispersion relation of capillary waves in a particular
approximation scheme. For simplicity, we focus on the simplest case in which
$\mathcal{Y}$ is a constant function of the temperature and chemical potential.
This implies that
$\mathcal{S}=\mathcal{Q}=\mathcal{X}_{TT}=\mathcal{X}_{T\mu}=\mathcal{X}_{\mu\mu}=0$,
hence Eqs.~\eqref{vanishingDet2} and~\eqref{vanishingDet3} are automatically
satisfied. In order to solve~\eqref{vanishingDet1} we focus on long-wavelength
perturbations so that $k_i\sim \tau $ and $\ell_a\sim \tau$ for a small
parameter $\tau$ and $k\ne0$.  In addition we consider the regime of small
charge $Q$ so that $Q,\chi_{T\mu},\chi_{\mu\mu}\sim\epsilon$ for a small
parameter $\epsilon$. In this situation \eqref{vanishingDet1} leads to
\begin{equation}
  \kappa = |k| - \frac{(k^2+\ell^2)}{2QS^2} \mathcal{Y}
  \left(\chi_{TT}Q-\chi_{T\mu}S\right)
  + \mathcal{O}\left(\tau^4, \epsilon\right)~~,\qquad
  \omega =\pm \sqrt{\frac{\kappa(k^2+\ell^2)\mathcal{Y}}{TS+\mathcal{E}\kappa}}
  + \mathcal{O}\left(\epsilon\right)~~,
\end{equation}
which in turn yields the dispersion relation
\begin{equation} \label{eq:disp}
  \omega = \sqrt{\frac{k(k^2+\ell^2)\mathcal{Y}}{ST+\mathcal{E}k}}
  \left(1 - (k^2+\ell^2) \mathcal{Y}T\frac{\left(\chi_{TT}Q -
        \chi_{T\mu}S\right)}{4kQS(\mathcal{E}k+ST)}
    + \mathcal{O}(k^{7/2},\epsilon)\right)~~.
\end{equation}
It is instructive to consider this expression in a particular limit. An interesting situation is the case in which
the perturbation only occurs along the $(p-q)$ transverse directions so that $\ell=0$.\footnote{The same scaling behaviour is observed if we take the perturbations to be the same in both directors, i.e.\ when $k^2=\ell^2$.} In this context, the
dispersion relation takes the form
\begin{equation} \label{eq:cap}
  \omega=k^{3/2}
  \sqrt{\frac{\mathcal{Y}}{ST}}\left(1-k\mathcal{Y}
    \frac{\left(\chi_{TT}Q - \chi_{T\mu}S\right)}{4QS^2}
    + \mathcal{O}(k^{7/2},\epsilon)\right)~~,
\end{equation}
where we have assumed that $k$ is small enough such that $\mathcal{E}k\ll
ST$. The leading $k^{3/2}$ behaviour is the classical result for the dispersion
relation for capillary waves of an uncharged fluid (see
e.g.~\cite{Armas:2016xxg}), while the sub-leading term of order $k^{5/2}$ is a
small deviation due to the presence of the dipole charge density $Q$.

Another interesting situation is one in which clear signatures of anisotropy are observed. Consider now the case in which there is 
no perturbation along the $(p-q)$ directions so that $k_i=0$. In the same regime where $\ell_a\sim \tau$ and $Q,\chi_{T\mu},\chi_{\mu\mu}\sim\epsilon$, \
Eq.~\eqref{vanishingDet1} now leads to
\begin{equation}
  \omega^2 = \frac{\ell^2 \mathcal{Y}\kappa}{TS+\mathcal{E}\kappa}+\mathcal{O}\left(\epsilon\right)~~,~~
  \kappa = - \ell^2\frac{\mathcal{Y}}{Q S^2}
  \left(\chi_{TT}Q-\chi_{T\mu}S\right)
  + \mathcal{O}(\tau^4,\epsilon)~~,
\end{equation}
so that the dispersion relation takes the form
\begin{equation} \label{eq:dispani}
\omega=\pm \ell^2 \sqrt{\frac{\mathcal{Y}^2}{TS^2 }\frac{"dd\log(Q/S)}{"dd T}}+\mathcal{O}\left(\tau^4,\epsilon\right)~~.
\end{equation}
This behaviour is a radical departure from the usual dispersion relation of capillary waves for isotropic fluids which behaves as~\eqref{eq:cap} and a clear signature of the presence 
of microscopic anisotropies. In order not to have unstable modes on the surface one must have that $\text{d}\log(Q/S)/dT>0$. If this condition does not hold, \eqref{eq:dispani} suggests that the surface would not form in any physical situation as a small perturbation along the $\Pi_{\mu\nu}$ directions would exponentially grow in time. In such cases, it would be interesting to understand the nature of the resulting surface instability. However, do note that this potential instability could be cured by considering the more physically relevant situation in which $\mathcal{Y}$ depends non-trivially on $T$ and $\mu$. In fact, in such context it would be interesting to study sound modes on the surface, analogously to \cite{Armas:2016xxg}. We leave this possibility for future work.

\section{Discussion}\label{sec:discussion}


In this paper we have introduced a framework for building effective theories of
hydrodynamics with higher-form symmetries. In particular, we have developed in
detail the case of fluids with a single $q$-form symmetry to first order in
derivatives. After defining the ideal order dynamics in Sec.~\ref{sec:ideal}, we
have found in Sec.~\ref{sec:diss} that the dissipative and parity-even sector of
the theory up to first order in derivatives is characterised by $q$+7
independent transport coefficients for $q>1$ and 7 for $q=1$, once Onsager's
relation is imposed. In comparison with the $q=1$ case
of~\cite{Grozdanov:2016tdf, Hernandez:2017mch}, there is one extra transport
coefficient for $q>1$ corresponding to the shear viscosity in the $\text{SO}(q)$
sector and $q-1$ extra current resistivities. In Sec.~\ref{sec:surface} we have
generalised these results in order to include the presence of an interface
separating distinct fluid phases and studied capillary waves on the interface,
which show signatures of anisotropy.  This analysis turns out to be similar to
the analysis carried out in the case of superfluids in~\cite{Armas:2016xxg}.

The work presented here suggests a few possible extensions:

\textbf{The hydrostatic sector:} In Sec.~\ref{sec:equilibrium} we have
constructed the most general equilibrium partition function under the assumption
of one timelike and $q$ spacelike isometries. We observed that this partition
function is more general than the one presented in~\cite{Grozdanov:2016tdf} for
$q=1$ and that it generalises for $q>1$ the solution provided by the free energy
obtained in~\cite{Caldarelli:2010xz} for $q=1$. Ref.~\cite{Grozdanov:2016tdf}
assumed that the isometries, in addition to having a vanishing Lie bracket, had
a vanishing inner product. This, however, is not necessarily the case as we have
explained.

Insisting on the existence of a hydrostatic sector for fluids with $q$-form
symmetry, regardless of the microscopic origin of the vectors $v_i$, we noticed
that equilibrium configurations do not exist unless constraints are imposed on
the background, in particular, the existence of $q$ spacelike commuting
isometries and an additional constraint, namely,
$"dd(\sigma\mathrm{Vol}_q - \imath_{\textbf{k}} A_{q+1}) =0$. This condition
guarantees that the equilibrium configurations obtained from~\eqref{eq:partition}
do not produce entropy. We understand that this is a peculiar feature of
fluids with higher-form symmetries as usually equilibrium only requires the
existence of a timelike isometry and no spatial isometries or extra
conditions. The need for these features originate from the fact that in
equilibrium one must satisfy the charge conservation equation
\begin{equation}
\nabla_\mu\left(Q Tv^\mu_1\right)=0~~,
\end{equation}
where we have specialised to the $q=1$ case for simplicity. A simple way to
satisfy this relation is to assume $v^\mu_1$ to be a linear combination of
background Killing vector fields at the expense of having to introduce the
ad-hoc requirement that
$"dd(\sigma\mathrm{Vol}_q - \imath_{\textbf{k}} A_{q+1}) =0$. On top of this, as
we commented in Sec.~\ref{newconstraints}, the equilibrium partition function seems to
impose new constraints on transport coefficients that do not follow from the
second law of thermodynamics. These restrictions are clearly unsatisfactory and
we intend to return to this issue in a forthcoming publication~\cite{toappear2}.

\textbf{Gravitational duals:} As mentioned in the introduction, one of the main
motivations of this work was to understand the structure of long-wavelength
perturbations of black branes in supergravity. In the context of the
fluid/gravity correspondence~\cite{Bhattacharyya:2008jc} and, more generally, in
the context of the blackfold approach~\cite{Emparan:2009cs, Emparan:2009at}, one
may test theories of hydrodynamics by appropriately perturbing certain classes
of black brane geometries. It is therefore interesting to consider gravitational
duals to these theories and perturb them in a derivative expansion. In the case
$q=1$ and in Anti-de Sitter space this was considered in
\cite{Grozdanov:2017kyl}. In a future publication~\cite{toappear1}, we consider
a more general class of black brane geometries valid for all $p$, $q$, $Q$ and
show that, for a constant external gauge field, their perturbations are characterised 
by the existence of 8 independent transport coefficients for $q>1$ in addition to the conservation equations for
the stress tensor and current appearing as constraint equations. In particular,
we show that this particular class of geometries satisfy the relations
\eqref{eq:etazeta} away from the isotropic limit. Since the second relation in
\eqref{eq:etazeta} is also satisfied for the class considered in
\cite{Grozdanov:2017kyl}, this suggests that the relations~\eqref{eq:etazeta}
may be universal for generic theories of gravity, at least those theories
without higher-derivative corrections.

\textbf{Fluids with multiple higher-form currents:} The case of fluids with multiple higher-form currents is of particular interest in the context of supergravity and string theory as generic black brane bound states in string theory may carry multiple higher-form charges. In particular, in the case of the D3-F1 bound state in type IIB string theory, the effective fluid carries two 2-form currents $j_2$ and $J_2$ dual to the NSNS two-form $B_2$ and the 2-form RR field, respectively, in addition to a 4-form current $J_4$ dual to the 4-form RR field~\cite{Armas:2016mes}. However, the current conservation equations are not trivial, instead they must satisfy
\begin{equation}
  \nabla_\mu j^{\mu\nu}_2=0~~,\qquad
  \nabla_\mu J^{\mu\mu_1\mu_2\mu_3}_4=0~~,\qquad
  \nabla_\mu J^{\mu\nu}_2=\frac{1}{3!}J_{4}^{\mu_1\mu_2\mu_3 \nu}H_{3\mu_1\mu_2\mu_3}~~,
\end{equation}
where $H_{3\mu_1\mu_2\mu_3}=\text{d}B_{2\mu_1\mu_2\mu_3}$. It would be interesting to address these systems in the future and to consider the most general type of hydrodynamic theory that can arise from supergravity~\cite{Armas:2016mes}. 

Another context where fluids with multiple higher-form currents have a role to play is in the context of the effective theory of charge density waves and states with dynamical defects~\cite{Andrade:2017cnc, Alberte:2017oqx, Amoretti:2017frz, Grozdanov:2018ewh}. The framework introduced in this work is capable of dealing with these cases and to provide a systematic construction of effective hydrodynamic theories with multiple higher-form symmetries.

\section*{Acknowledgements}
We would like to thank J. Bhattacharya and D. Hofman for various helpful
discussions. JA would like to thank R. Emparan for a motivational discussion on
this subject. AJ would like to thank Perimeter Institute,
where part of this project was done, for hospitality. JA is partly supported by the Netherlands
Organization for Scientific Research (NWO). AJ is supported by the Durham
Doctoral Scholarship offered by Durham University. The work of AVP has been supported by The Danish Council for Independent Research - Natural Sciences (FNU), DFF-4002-00307. 
AVP gratefully acknowledges support from UC Berkeley where some of the research for this paper was carried out.

\appendix

\section{Currents in another fluid frame}\label{app:frames}
In this appendix we briefly compare the transport coefficients in the frame chosen in Sec.~\ref{sec:frames} with those presented in~\cite{Grozdanov:2016tdf} for $q=1$. The two frames differ from each other due to the transport coefficient $\Theta$ in the current~\eqref{eq:1storder}. Using the frame transformation~\eqref{eq:framet} in order to remove this term from the current we require
\begin{equation}
\overline\delta v_1^\mu = 2 \Theta\Gamma^{\mu\nu}v_1^\lambda\nabla_{(\nu}u_{\lambda)}~~,
\end{equation}
which due to~\eqref{eq:framet} adds an extra term to the stress tensor with one index along $v_1^\mu$ and another along the normal directions $\overline\delta v_1^\mu$. Comparing with Eqs.~(3.9)-(3.14) of~\cite{Grozdanov:2016tdf} and using~\eqref{eq:1storder} we identify the transport coefficients
\begin{equation}
\eta_{||}=Q \mu\Theta~~,~~\eta_\perp=\eta_\Gamma~~,~~r_{\perp}=\mathfrak{D}T~~,~~r_{||}=\mu\Omega_1~~,~~\zeta_{||}=\zeta_\Pi~~,~~\zeta_\perp=\zeta_\Gamma~~,~~\zeta_\times=\zeta~~,
\end{equation}
where the coefficients $\eta_{||}$, $\eta_\perp$, $r_{\perp}$, $r_{||}$,
$\zeta_{||}$, $\zeta_\perp$ and $\zeta_\times$ were introduced
in~\cite{Grozdanov:2016tdf}.


\addcontentsline{toc}{section}{References}
\footnotesize
\providecommand{\href}[2]{#2}\begingroup\raggedright\endgroup
\end{document}